\documentclass[prd,amsfonts,noshowpacs,nofootinbib]{revtex4}
\usepackage{amsmath}

\newcommand{\B}[1]{\mathbf{#1}}

\begin{document}

\begin{abstract}
We compute the fourth order action in perturbation theory for
scalar and second order tensor perturbations for a minimally
coupled single field inflationary model, where the inflaton's
lagrangian is a general function of the field's value and its
kinetic energy. We obtain the fourth order action
in two gauges, the comoving gauge and the uniform curvature gauge.
Using the comoving gauge action we calculate the trispectrum at
leading order in slow roll, finding agreement with a previously
known result in the literature. We point out that in general to
obtain the correct leading order trispectrum one cannot ignore
second order tensor perturbations as previously done by others.
The next-to-leading order corrections may become detectable
depending on the shape and we provide the necessary formalism to
calculate them.
\end{abstract}

\title{Non-gaussianity from the trispectrum in general single field inflation}
\author{Frederico Arroja\footnote{Frederico.Arroja@port.ac.uk}}
\author{Kazuya Koyama\footnote{Kazuya.Koyama@port.ac.uk}}
\affiliation{Institute of Cosmology and Gravitation, University of
Portsmouth, Portsmouth PO1 2EG, UK}
\date{\today}
\maketitle

\section{\label{sec:INTRO}Introduction}

The theory of slow-roll inflation generically predicts that the
observed cosmic microwave background radiation (CMBR) anisotropies
are nearly scale invariant and very gaussian. Indeed, the latest
observations of CMBR by WMAP3 \cite{Spergel:2006hy} confirm these
expectations. This constitutes one of the biggest achievements of
modern cosmology.

Despite its successes the theory of inflation still has many open
questions. For example, we do not know the origin of the scalar
field whose energy drives inflation, not to mention that we have
never detected directly in the laboratory these kind of particles.
The energy scale at which inflation happened is unknown by many
orders of magnitude. There are many models of inflation that give
similar predictions for the power spectrum of primordial
perturbations, so which one (if any) is the correct one?

For us to move a step forward in our understanding of the very
early universe we have to work in two fronts. First the
observational side. In the next few years, with improved
experiments like the Planck satellite, we will measure the CMBR
anisotropies to an incredible resolution. For example the
observational bounds on the bispectrum (the three point
correlation function of the primordial curvature perturbation
$\zeta$) will shrink from the present WMAP3 value $-50<f_{NL}<114$
\cite{Spergel:2006hy} to $|f_{NL}|\lesssim 5$
\cite{Komatsu:2001rj}, where the parameter $f_{NL}$ parameterizes
the size of the bispectrum. It's because this parameter is
constrained to be small that we say that the CMBR anisotropies are
very gaussian. The observational bounds on the trispectrum (four
point function) will also tighten significantly from the rather
weak present constraint of $|\tau_{NL}|<10^8$
\cite{Boubekeur:2005fj,Alabidi:2005qi} to the future constraint of
$|\tau_{NL}|\sim 560$ \cite{Kogo:2006kh}, where $\tau_{NL}$
denotes the size of the trispectrum. These previous observational
bounds on the non-linearity parameters are for non-gaussianity of
the local type. These bounds change depending on the shape of the
wave vectors' configuration \cite{Babich:2004gb}. This is one of
the reasons why it is important to calculate the shape dependence
of the non-gaussianity.

In face of these expected observational advances, it is then
imperative to push forward our theoretical knowledge of our
theories and calculate more observational consequences of the
different inflationary models to make a comparison with
observations possible. One possible direction to be followed by us
and many others is to calculate higher order statistics (like the
trispectrum) of the primordial curvature perturbation. These
higher order statistics contain much more information about the
inflationary dynamics and if we observe them they will strongly
constrain our models. Because these higher order statistics have a
non-trivial momentum dependence (shape) they will help to
discriminate between models that have a similar power spectrum
(two point function).

Calculations of the bispectrum for a single field inflationary
model were done by Maldacena \cite{Maldacena:2002vr}. He showed
that the primordial bispectrum is too small (of the order of the
slow-roll parameters) to be observed even with Planck. Subsequent
work generalized Maldacena's result to include more fields and
more complicated kinetic terms \cite{Seery:2005wm, Seery:2005gb,
Chen:2006nt}. In \cite{Chen:2006nt}, Chen \emph{et al.} have
calculated the bispectrum for a quite general model of single
field inflation. They showed that for some models even the
next-to-leading order corrections in the slow-roll expansion may
be observed.

In this paper, we will focus our attention on the calculation of
the trispectrum. In \cite{Seery:2006vu}, Seery \emph{et al.} have
calculated the trispectrum for slow-roll multi-field models (with
standard kinetic terms) and they showed that at horizon crossing
it is too small to be observed. But there are models of single
field inflation, well motivated from more fundamental theories,
that can produce a significant amount of non-gaussianity, such as
Dirac-Born-Infeld (DBI) inflation
\cite{Alishahiha:2004eh,Huang:2006eh}. In \cite{Huang:2006eh} the
authors have computed the trispectrum for a model where the
inflaton's lagrangian is a general function of the field's kinetic
energy and the field's value, their result was obtained using a
simple method \cite{Gruzinov:2004jx}, that only gives the correct
leading order answer. In this paper we will provide the equations
necessary to calculate the next-to-leading order corrections to
the trispectrum. We argue that for some models these corrections
might become equally observable in the future. In fact, we will
calculate the fourth order action in the uniform curvature gauge
that is exact in the slow-roll expansion and therefore in
principle one could calculate all slow-roll corrections to the
trispectrum of the field perturbations.

We will also compute the exact fourth order action for the
curvature perturbation $\zeta$ in the comoving gauge. For a
simpler inflation model (with the standard kinetic term) this was
recently done in \cite{Jarnhus:2007ia}. However
\cite{Jarnhus:2007ia} did not consider second order tensor
perturbations. We will argue that this is an oversimplification
that leads to erroneous results. The reason simply being that at
second order in perturbation theory, scalar degrees of freedom
will source second order tensor perturbations and this will give a
non-zero contribution for the fourth order action and hence the
trispectrum.

There are other reasons why we will perform the calculation in the
comoving gauge. First of all, in doing so we work all the time
with the gauge invariant variable $\zeta$ that is directly related
with the observational quantities. The comoving gauge action can
also be used other practical purposes. For example, it can be used
to calculate loop effects that can possibly have important
observational consequences. It can also be used to calculate the
trispectrum of models where the potential has a ``feature" (see
\cite{Chen:2006xjb,Chen:2008wn} for an example of such calculation
for the bispectrum). In the vicinity of the sudden potential
``jump" the slow-roll approximation temporarily fails and one
might get an enhancement of the trispectrum. There are well
motivated models of brane inflation where the throat's warp factor
suddenly jumps \cite{Hailu:2006uj}.

This paper is organized as follows. In the next section, we
introduce the model under consideration. In section
\ref{sec:Perturbations} we shall study non-linear perturbations.
First, we compute the fourth order action in the comoving gauge
including both scalar and second order tensor degrees of freedom.
After that we compute the fourth order action in the uniform
curvature gauge. In section \ref{sec:trispectrumformalism}, we
present the formalism needed to calculate the trispectrum. In
section \ref{sec:leadingtrispectrum}, we calculate the leading
order trispectrum using the comoving gauge action. We comment on
previous works and on the observability of next-to-leading order
corrections. Section \ref{sec:conclusion} is devoted to
conclusions. Finally in Appendix, we present the second order
gauge transformation between the two previous gauges and a way to
extract the transverse and traceless part of a tensor.

\section{\label{sec:MODEL}The model}
In this work, we will consider a fairly general class of models
described by the following action
\begin{equation}
S=\frac{1}{2}\int d^4x\sqrt{-g}\left[M^2_{Pl}R+2P(X,\phi)\right],
\label{action}
\end{equation}
where $\phi$ is the inflaton field, $M_{Pl}$ is the Planck mass
that we will set to unity hereafter, $R$  is the Ricci scalar and
$X\equiv-\frac{1}{2}g^{\mu\nu}\partial_\mu\phi\partial_\nu\phi$ is
the inflaton's kinetic energy. $g_{\mu\nu}$ is the metric tensor.
We label the inflaton's lagrangian by $P$ and we assume that it is
a well behaved function of two variables, the inflaton field and
its kinetic energy.

This general field lagrangian includes as particular cases the
common slow-roll inflation model, DBI-inflation
\cite{Silverstein:2003hf} \cite{Alishahiha:2004eh} and K-inflation
\cite{ArmendarizPicon:1999rj}.

We are interested in flat, homogeneous and isotropic
Friedmann-Robertson-Walker universes described by the line element
\begin{equation}
ds^2=-dt^2+a^2(t)\delta_{ij}dx^idx^j, \label{FRW}
\end{equation}
where $a(t)$ is the scale factor. The Friedmann equation and the
continuity equation read
\begin{equation}
3H^2=E, \label{EinsteinEq}
\end{equation}
\begin{equation}
\dot{E}=-3H\left(E+P\right), \label{continuity}
\end{equation}
where the Hubble rate is $H=\dot{a}/a$, $E$ is the energy of the
inflaton and it is given by
\begin{equation}
E=2XP_{,X}-P
,\label{energy}
\end{equation}
where $P_{,X}$ denotes the derivative of $P$ with respect to $X$.

It was shown in \cite{Garriga:1999vw} that for this model the
speed of propagation of scalar perturbations (``speed of sound")
is $c_s$ given by
\begin{equation}
c_s^2=\frac{P_{,X}}{E_{,X}}=\frac{P_{,X}}{P_{,X}+2XP_{,XX}}. \label{sound
speed}
\end{equation}
We define the slow variation parameters, analogues of the
slow-roll parameters, as:
\begin{equation}
\epsilon=-\frac{\dot{H}}{H^2}=\frac{XP_{,X}}{H^2}, \label{epsilon}
\end{equation}
\begin{equation}
\eta=\frac{\dot{\epsilon}}{\epsilon H}, \label{eta}
\end{equation}
\begin{equation}
s=\frac{\dot{c_s}}{c_sH}. \label{s}
\end{equation}
We should note that these slow variation parameters are more
general than the usual slow-roll parameters and that the smallness
of these parameters does not imply that the field in rolling
slowly. We assume that the rate of change of the speed of sound is
small (as described by $s$) but $c_s$ is otherwise free to change
between zero and one.

It is convenient to introduce the following parameters that
describe the non-linear dependence of the lagrangian on the
kinetic energy:
\begin{equation}
\Sigma=XP_{,X}+2X^2P_{,XX}=\frac{H^2\epsilon}{c_s^2},
\label{Sigma}
\end{equation}
\begin{equation}
\lambda=X^2P_{,XX}+\frac{2}{3}X^3P_{,XXX}, \label{lambda}
\end{equation}
\begin{equation}
\Pi=X^3P_{,XXX}+\frac{2}{5}X^4P_{,XXXX}. \label{Pi}
\end{equation}
These parameters are related to the size of the bispectrum and
trispectrum. The power spectrum of the primordial quantum
fluctuation was first derived in \cite{Garriga:1999vw} and reads
\begin{equation}
P_k^\zeta=\frac{1}{36\pi^2}\frac{E^2}{E+P}=\frac{1}{8\pi^2}\frac{H^2}{c_s\epsilon},
\label{PowerSpectrum}
\end{equation}
where it should be evaluated at the time of horizon crossing
${c_s}_*k=a_*H_*$. The spectral index is
\begin{equation}
n_s-1=\frac{d\ln P_k^\zeta}{d\ln k}=-2\epsilon-\eta-s.
\label{SpectralIndex}
\end{equation}
WMAP observations of the perturbations in the CMBR tell us that
the previous power spectrum is almost scale invariant therefore
implying that the three slow variation parameters should be small
at horizon crossing, roughly of order $10^{-2}$.

\section{\label{sec:Perturbations}Non-linear
perturbations}

In this section, we will consider perturbations of the background
(\ref{FRW}) beyond linear order. There is a vast literature on
second order perturbations that are important when one is
interested in calculating three point correlation functions, see
for example
\cite{Acquaviva:2002ud,Maldacena:2002vr,Seery:2005wm,Chen:2006nt}.
In the present paper, we are interested in non-gaussianities that
come from the trispectrum and so we need to use third order
perturbation theory. For that we need to compute the fourth order
in the perturbation action. In this section we will obtain the
fourth order action in two different gauges. As a check on our
calculations we will compute the leading order (in slow roll)
trispectrum in both gauges. We will follow the pioneering approach
developed by Maldacena \cite{Maldacena:2002vr} and used in several
subsequent papers
\cite{Seery:2005wm,Seery:2005gb,Huang:2006eh,Seery:2006vu}.

For reasons that will become clear later it is convenient to use
the ADM metric formalism \cite{Arnowitt:1960es}. The ADM line
element reads
\begin{equation}
ds^2=-N^2dt^2+h_{ij}\left(dx^i+N^idt\right)\left(dx^j+N^jdt\right),
\label{ADMmetricphi}
\end{equation}
where $N$ is the lapse function, $N^i$ is the shift vector and
$h_{ij}$ is the 3D metric.

The action (\ref{action}) becomes
\begin{equation}
S=\frac{1}{2}\int dtd^3x\sqrt{h}N\left({}^{(3)}\!R+2P\right)+
\frac{1}{2}\int dtd^3x\sqrt{h}N^{-1}\left(E_{ij}E^{ij}-E^2\right).
\end{equation}
The tensor $E_{ij}$ is defined as
\begin{equation}
E_{ij}=\frac{1}{2}\left(\dot{h}_{ij}-\nabla_iN_j-\nabla_jN_i\right),
\end{equation}
and it is related to the extrinsic curvature by
$K_{ij}=N^{-1}E_{ij}$. $\nabla_i$ is the covariant derivative with
respect to $h_{ij}$ and all contra-variant indices in this section
are raised with $h_{ij}$ unless stated otherwise.

The hamiltonian and momentum constraints are respectively
\begin{eqnarray}
{}^{(3)}\!R+2P-2\pi^2N^{-2}P_{,X}-N^{-2}\left(E_{ij}E^{ij}-E^2\right)&=&0,\nonumber\\
\nabla_j\left(N^{-1}E_i^j\right)-\nabla_i\left(N^{-1}E\right)&=&\pi
N^{-1}\nabla_i\phi P_{,X},\label{LMphi}
\end{eqnarray}
where $\pi$ is defined as
\begin{equation}
\pi\equiv \dot{\phi}-N^j\nabla_j\phi.\label{pi}
\end{equation}
We decompose the shift vector $N^i$ into scalar and intrinsic
vector parts as
\begin{equation}
N_i=\tilde{N_i}+\partial_i\psi,
\end{equation}
where $\partial_i\tilde{N^i}=0$, here indices are raised with
$\delta_{ij}$.

Before we consider perturbations around our background let us
count the number of degrees of freedom (dof) that we have. There
are five scalar functions, the field $\phi$, $N$, $\psi$,
$\mbox{det} h$ and $h_{ij}\sim\partial_i\partial_j H$, where $H$
is a scalar function and $\mbox{det} h$ denotes the determinant of
the 3D metric. Also, there are two vector modes $\tilde{N}^i$ and
$h_{ij}\sim\partial_i\chi_j$, where $\chi^j$ is an arbitrary
vector. Both $\tilde{N}^i$ and $\chi^j$ satisfy a divergenceless
condition and so carry four dof. Furthermore, we also have a
transverse and traceless tensor mode $\gamma_{ij}$ that contains
two additional dof. Because our theory is invariant under change
of coordinates we can eliminate some of these dof. For instance, a
spatial reparametrization like
$x^i=\tilde{x}^i+\partial^i\tilde{\epsilon}(\tilde{x},\tilde{t})+\epsilon^i_{(t)}(\tilde{x},\tilde{t})$,
where $\tilde{\epsilon}$ and  $\epsilon^i_{(t)}$ are arbitrary and
$\partial_i\epsilon^i_{(t)}=0$, can be chosen so that it removes
one scalar dof and one vector mode. A time reparametrization would
eliminate another scalar dof. Constraints in the action will
eliminate further two scalar dof and a vector mode. In the end we
are left with one scalar, zero vector and one tensor modes that
correspond to three physical propagating dof.

In the next subsection we shall use two different gauges that
correctly parameterize these dof. Because physical observables are
gauge invariant we know that both gauges have to give the same
result for the trispectrum for instance. It seems then unnecessary
to perform the calculation twice in different gauges. In practice,
we will see that both gauges have advantages and disadvantages and
one is more suitable for some applications than the other.
Furthermore, it provides a good consistency check on the
calculation.

\subsection{\label{subsec:PerturbationsComoving}Non-linear
perturbations in the comoving gauge}

In this subsection, we will compute the fourth order action for
the general model (\ref{action}) in the comoving gauge. In this
gauge the scalar degree of freedom is the so-called curvature
perturbation $\zeta$ that is also gauge invariant. There are a few
works on this subject using this gauge, see e.g.
\cite{Jarnhus:2007ia}, where the authors have calculated the
fourth order action for a standard kinetic term inflation but they
neglected second order tensor perturbations. We will show that
this is an oversimplification that may lead to an erroneous result
for the four point correlation function.

In the comoving gauge, the inflaton fluctuations vanish and the 3D
metric is perturbed as
\begin{eqnarray}
&&\delta\phi=0,\nonumber\\  &&h_{ij}=a^2e^{2\zeta}\hat{h}_{ij},
\quad
\hat{h}_{ij}=\delta_{ij}+\gamma_{ij}+\frac{1}{2}\gamma_{ik}\gamma_j^k+\cdots\label{zetagauge}
\end{eqnarray}
where $\mbox{det} \hat{h}=1$, $\gamma_{ij}$ is a tensor
perturbation that we assume to be a second order quantity, i. e.
$\gamma_{ij}=\mathcal{O}(\zeta^2)$. It obeys the traceless and
transverse conditions $\gamma_i^i=\partial^i\gamma_{ij}=0$
(indices are raised with $\delta_{ij}$). $\zeta$ is the gauge
invariant scalar perturbation. In (\ref{zetagauge}), we have
ignored the first order tensor perturbations
${}^{(1)}{\gamma_{ij}}_{GW}$. This is because any correlation
function involving this tensor mode will be smaller than a
correlation function involving only scalars, see results of
\cite{Maldacena:2002vr}. In the literature the second order tensor
perturbations are often neglected, however they should be taken
into account. The reason for this is because at second order the
scalars will source the tensor perturbations equation. Later in
this section, we will elaborate further on this point. Higher
order tensor perturbations, like ${}^{(3)}\gamma_{ij}$, do not
contribute to the fourth order action.

We expand $N$ and $N^i$ in power of the perturbation $\zeta$
\begin{eqnarray}
N=1+\alpha_1+\alpha_2+\cdots,\\
\tilde{N_i}=\tilde{N_i}^{(1)}+\tilde{N_i}^{(2)}+\cdots,\\
\psi=\psi_1+\psi_2+\cdots,
\end{eqnarray}
where $\alpha_n$, $\tilde{N_i}^{(n)}$ and $\psi_n$ are of order
$\zeta^n$.

Some useful expressions for the quantities appearing in
(\ref{LMphi}), valid to all orders in perturbations but for
$\gamma_{ij}=0$:
\begin{eqnarray}
{}^{(3)}\!R=-2a^{-2}e^{-2\zeta}\left(\partial_i\zeta\partial^i\zeta+2\partial_i\partial^i\zeta\right),
\end{eqnarray}
\begin{eqnarray}
E_{ij}E^{ij}-E^2&=&-6\left(H+\dot{\zeta}\right)^2+\frac{4H}{a^2}\left(1+\frac{\dot{\zeta}}{H}\right)e^{-2\zeta}
\left(\partial^2\psi+\partial^k\zeta\partial_k\psi+\partial^k\zeta\tilde{N}_k\right)\nonumber
\\\nonumber&&
+a^{-4}e^{-4\zeta}\Big[\frac{1}{2}\partial_i\tilde{N}_j\left(\partial^i\tilde{N}^j+\partial^j\tilde{N}^i\right)+2\partial^i\partial^j\psi\partial_i\tilde{N}_j
+\partial_i\partial_j\psi\partial^i\partial^j\psi-\partial^2\psi\partial^2\psi
\\\nonumber&&
-2\partial_i\tilde{N}_j\left(\partial^j\zeta\tilde{N}^i+\partial^i\zeta\tilde{N}^j\right)-4\tilde{N}_k\partial_i\zeta\partial^i\partial^k\psi
-2\partial_i\tilde{N}_j\left(\partial^j\zeta\partial^i\psi+\partial^i\zeta\partial^j\psi\right)
\\&&
-4\partial_i\partial_j\psi\partial^j\zeta\partial^i\psi+2\partial^j\zeta\left(\partial_j\zeta\tilde{N}_i\tilde{N}^i+2\partial_j\zeta\tilde{N}_i\partial^i\psi+\partial_j\zeta\partial_i\psi\partial^i\psi\right)
\Big],
\end{eqnarray}
\begin{eqnarray}
&&\nabla_j\left(N^{-1}E_i^j\right)-\nabla_i\left(N^{-1}E\right)=
\nonumber\\ && -N^{-2}\Big[-2\left(H+\dot{\zeta}\right)\partial_iN
+\frac{a^{-2}e^{-2\zeta}}{2}\left(-\partial^jN\left(\partial_iN_j+\partial_jN_i\right)+2\partial_jN\left(\partial^j\zeta
N_i+\partial_i\zeta
N^j\right)+2\partial_iN\partial^2\psi\right)\Big] \nonumber\\ &&
+N^{-1}\Big[-2\partial_i\dot{\zeta}
+a^{-2}e^{-2\zeta}\left(\frac{1}{2}\left(\partial^j\zeta\partial_j\tilde{N}_i-\partial_j\zeta\partial_i\tilde{N}^j\right)
+\partial_j\zeta\partial^j\zeta N_i-\partial_i\zeta\partial^j\zeta
N_j+\partial_i\partial^j\zeta N_j+\partial^2\zeta
N_i-\frac{1}{2}\partial^2\tilde{N}_i\right)\Big].\nonumber\\
\end{eqnarray}
In the previous equations, indices in the right-hand side are
raised with $\delta_{ij}$ while indices in the left-hand side are
raised with $h_{ij}$. In the rest of this section indices will be
raised with $\delta_{ij}$.

Now, the strategy is to solve the constraint equations for the
lapse function and shift vector in terms of $\zeta$ and then plug
in the solutions in the expanded action up to fourth order.

At first order in $\zeta$, a particular solution for equations
(\ref{LMphi}) is \cite{Maldacena:2002vr,Seery:2005wm}:
\begin{equation}
\alpha_1=\frac{\dot{\zeta}}{H}, \quad \tilde{N_i}^{(1)}=0, \quad
\psi_1=-\frac{\zeta}{H}+\chi, \quad
\partial^2\chi=a^2\frac{\epsilon}{c_s^2}\dot{\zeta}.\label{N1order}
\end{equation}
At second order, the constraint equation for the lapse function
gives
\begin{eqnarray}
\frac{4H}{a^2}\partial^2\psi_2&=&-2a^{-2}\partial_i\zeta\left(\partial^i\zeta+2H\partial^i\psi_1\right)
-4\alpha_1\left(a^{-2}\partial_i\partial^i\zeta-2\Sigma\zeta\right)-2\alpha_1^2\left(\Sigma+6\lambda\right)\nonumber\\
&&-a^{-4}\left(\partial^i\partial_k\psi_1\partial_i\partial^k\psi_1-\partial^2\psi_1\partial^2\psi_1\right)
+4\alpha_2\left(\Sigma-3H^2\right),
\end{eqnarray}
and the equation for the shift vector gives
\begin{equation}
2H\partial_i\alpha_2-\frac{1}{2}a^{-2}\partial^2\tilde{N_i}^{\!\!(2)}=
-a^{-2}\left(\partial_k\alpha_1\partial^k\partial_i\psi_1-\partial_i\alpha_1\partial^2\psi_1
+\partial^2\zeta\partial_i\psi_1+\partial_i\partial^k\zeta\partial_k\psi_1\right).
\label{Niconstraint2order}
\end{equation}
Due to the fact that $\tilde{N^i}$ is divergenceless and that any
vector can be separated into a incompressible and irrotational
part one can separate the contributions from $\alpha_2$ and
$\tilde{N_i}^{\!\!(2)}$ in the previous equation. The irrotational
part of Eq. (\ref{Niconstraint2order}) gives
\begin{equation}
2H\alpha_2=\partial^{-2}\partial^iF_i, \label{alpha2}
\end{equation}
and the incompressible part gives
\begin{equation}
\frac{1}{2a^2}\tilde{N_i}^{\!\!(2)}=-\partial^{-2}F_i+
\partial^{-4}\partial_i\partial^kF_k, \label{Ni2}
\end{equation}
where $F_i$ is define as the right-hand side of equation
(\ref{Niconstraint2order}). The operator $\partial^{-2}$ is
defined by $\partial^{-2}(\partial^2\varphi)=\varphi$ and in
Fourier space it just bring in a factor of $-1/k^2$.

It was shown in \cite{Chen:2006nt} that to compute the effective
action of order $\zeta^n$, within the ADM formalism, one only
needs to use the solution for the Lagrange multipliers $N$ and
$N^i$ up to order $\zeta^{n-2}$. Therefore in order to calculate
the fourth order effective action the knowledge of the Lagrange
multipliers up to second order is required. It is given in
equations (\ref{N1order}), (\ref{alpha2}) and (\ref{Ni2}).

The second order action is
\begin{equation}
S_2=\int
dtd^3x\left[a^3\frac{\epsilon}{c_s^2}\dot{\zeta}^2-a\epsilon\left(\partial\zeta\right)^2\right].
\label{2action}
\end{equation}

The third order action is
\cite{Maldacena:2002vr,Seery:2005wm,Chen:2006nt}
\begin{eqnarray}
S_3&=&\int dtd^3x\left[-\epsilon
a\zeta\left(\partial\zeta\right)^2-a^3\left(\Sigma+2\lambda\right)\frac{\dot{\zeta}^3}{H^3}
+\frac{3a^3\epsilon}{c_s^2}\zeta\dot{\zeta}^2\right.\nonumber\\
&&\left.+\frac{1}{2a}\left(3\zeta-\frac{\dot{\zeta}}{H}\right)\left(\partial_i\partial_j\psi_1\partial_i\partial_j\psi_1-\partial^2\psi_1\partial^2\psi_1\right)
-\frac{2}{a}\partial_i\psi_1\partial_i\zeta\partial^2\psi_1\right].\label{3action}
\end{eqnarray}

The scalar fourth order action is
\begin{eqnarray}
S_4&=&\frac{1}{2}\int dtd^3xa^3\left[-a^{-2}\epsilon
\zeta^2\left(\partial\zeta\right)^2+\alpha_1^4\left(2\Sigma+9\lambda+\frac{10}{3}\Pi\right)
-6\zeta\alpha_1^3\left(\Sigma+2\lambda\right)+9\zeta^2\alpha_1^2\Sigma\right.\nonumber\\
&&-2\alpha_2^2\left(\Sigma-3H^2\right)
+a^{-4}\left(\frac{\zeta^2}{2}+\zeta\alpha_1+\alpha_1^2\right)\left(\partial^kN_j^{(1)}\partial_k{N^j}^{(1)}-\partial^jN_j^{(1)}\partial^kN_k^{(1)}\right)\nonumber\\
&&-2a^{-4}\left(\zeta+\alpha_1\right)\partial^k{N^j}^{(1)}\left(\partial_kN_j^{(2)}-\delta_{kj}\partial^nN_n^{(2)}-2\partial_j\zeta N_k^{(1)}\right)\nonumber\\
&&+a^{-4}\left(-4\partial_kN_j^{(1)}\partial^j\zeta
{N^k}^{(2)}+2N_k^{(1)}\partial_j\zeta {N^k}^{(1)}\partial^j\zeta\right)\nonumber\\
&&\left.+\frac{a^{-4}}{2}\partial_k\tilde{N}_j^{(2)}\partial^k\tilde{N}^{j(2)}-2a^{-4}\partial_kN_j^{(2)}\left(\partial^j\zeta
{N^k}^{(1)}+\partial^k\zeta {N^j}^{(1)}\right)\right].
\label{SSSS}
\end{eqnarray}
Here, no slow-roll approximation has been made. The previous
action has to be supplemented with the action containing terms
with one and two tensors
\begin{equation}
S_{\gamma^2}=\frac{1}{8}\int
dtd^3x\left[a^3\dot{\gamma}_{ij}\dot{\gamma}^{ij}-a\partial_k\gamma_{ij}\partial^k\gamma^{ij}\right],\label{GG}
\end{equation}
\begin{equation}
S_{\gamma\zeta^2}=\int
dtd^3x\left[-2\frac{a}{H}\gamma_{ij}\partial^i\dot{\zeta}\partial^j\zeta-a\gamma_{ij}\partial^i\zeta\partial^j\zeta
-\frac{1}{2}a\left(3\zeta-\frac{\dot{\zeta}}{H}\right)\dot{\gamma}_{ij}\partial^i\partial^j\psi_1
+\frac{1}{2}a^{-1}\partial_k\gamma_{ij}\partial^i\partial^j\psi_1\partial^k\psi_1\right].\label{GSS}
\end{equation}

\subsubsection{\label{subsubsec:MSvariable}Canonical variable for quantization $\zeta_n$}

In order to calculate the quantum four point correlation function
we follow the standard procedure in quantum field theory. However
there is an important subtlety here. The gauge invariant quantity
$\zeta$ is not the correct variable to be quantized, because it is
not a canonical field . The canonical field to be quantized is the
field perturbation $\delta\phi$, or a convenient parameterization
$\zeta_n$ defined by
\begin{equation}
\zeta_n=-\frac{H}{\dot{\phi}_0}\delta\phi,\label{zetan}
\end{equation}
where $\phi_0$ is the background value of the field. We will see
that $\zeta$ is related to $\zeta_n$ by a non-linear
transformation so for the power spectrum calculation both
procedures of quantizing $\zeta$ or $\zeta_n$ give the same answer
because the difference between these two variables is a second
order quantity. However, for the calculation of higher order
correlation functions (like the bispectrum or trispectrum)
$\zeta_n$ is the correct variable to be quantized as it is linear
in $\delta\phi$. For the trispectrum, quantizing $\zeta$ gives
different results from quantizing $\zeta_n$. We will find the
relation between $\zeta$ and $\zeta_n$ through the gauge
transformation equations from the uniform curvature gauge
(discussed in detail in the next subsection) to the comoving
gauge. In the uniform curvature gauge the ansatz is
\begin{eqnarray}
&&\phi(x,t)=\phi_0+\delta\phi(x,t),\nonumber\\
&&h_{ij}=a^2\hat{h}_{ij}, \quad
\hat{h}_{ij}=\delta_{ij}+\tilde{\gamma}_{ij}+\frac{1}{2}\tilde{\gamma}_{ik}\tilde{\gamma}_j^k+\cdots\label{deltaphigaugeA}
\end{eqnarray}
where $\mbox{det} \hat{h}=1$ and $\tilde{\gamma}_{ij}$ is a tensor
perturbation that we assume to be a second order quantity, i.e.,
$\tilde{\gamma}_{ij}=\mathcal{O}(\delta\phi^2)$. The gauge
transformations are
\begin{eqnarray}
\zeta(\B{x})&=&\zeta_n(\B{x})+F_2(\zeta_n(\B{x}))+F_3(\zeta_n(\B{x})),\label{gte0}\\
\gamma_{ij}&=&\tilde{\gamma}_{ij}(t)+\mu_{ij}\label{gtt},
\end{eqnarray}
where $F_2(\zeta_n),\mu_{ij}=\mathcal{O}(\zeta_n^2)$,
$F_3(\zeta_n)=\mathcal{O}(\zeta_n^3)$ are the terms coming from
the second and third order gauge transformations respectively,
they can be found explicitly in Appendix \ref{GT} or in
\cite{Maldacena:2002vr,Jarnhus:2007ia}.

We now need to find the fourth order action for the variable
$\zeta_n$. Schematically, we write the different contributions as
\begin{equation}
S_{\zeta_n}=S_4(\zeta_n)+S_{\gamma^2}(\zeta_n)+S_{\gamma\zeta^2}(\zeta_n)+S_3(F_2(\zeta_n))+S_2(F_2(\zeta_n)).\label{zetanaction}
\end{equation}
The first three terms come from Eqs. (\ref{SSSS}-\ref{GSS}) when
we substitute $\zeta$ with $\zeta_n$ and $\gamma_{ij}$ with Eq.
(\ref{gtt}). Due to the non-linear relation between $\zeta$ and
$\zeta_n$, the third order action for $\zeta$, Eq.
(\ref{3action}), will after the change of variables give a
contribution to the fourth order action like $S_3(F_2(\zeta_n))$.
Similarly, the second order action, Eq. (\ref{2action}), will also
contribute with $S_2(F_2(\zeta_n))$. In principle, one would also
need to compute the third order gauge transformation as the second
order action gives origin to fourth order terms like
$\dot{\zeta_n}\dot{F}_3(\zeta_n)$ (where $F_3(\zeta_n)$ is the
third order piece of the gauge transformation). Fortunately, terms
involving $F_3$ can be shown to be proportional to the first order
equations of motion for $\zeta_n$, therefore when computing the
trispectrum these terms will vanish and we don't need to calculate
the third order gauge transformation explicitly at this point. It
can be easily seen that equations (\ref{3action}), (\ref{SSSS}),
(\ref{GG}), (\ref{GSS}) or their counterparts in terms of
$\zeta_n$, Eq. (\ref{zetanaction}), have terms that are not slow
roll suppressed. However, because in pure de Sitter space $\zeta$
is a gauge mode we expect the action (\ref{zetanaction}) to be
slow-roll suppressed (of order $\epsilon$). One can perform many
integrations by parts to show that the unsuppressed terms of
(\ref{zetanaction}) can be reduced to total derivative terms and
slow-roll suppressed terms given by
\begin{eqnarray}
S_{parts}=\int dt&&\!\!\!\!\!\!d^3x\Big\{
-\frac{3\epsilon}{aH^2}\Big[
\frac{1}{4}\left(\partial_j\zeta_n\partial^j\zeta_n\right)^2+\frac{1}{8}\zeta_n^2\partial^2\left(\partial_j\zeta_n\partial^j\zeta_n\right)
+\partial_j\zeta_n\partial^j\zeta_n\partial^{-2}\partial^l\partial^k\left(\partial_l\zeta_n\partial_k\zeta_n\right)
\nonumber
\\&&\;\;\;\;\;\;\;\;\;\;\;\;\;\;\;\;\;\;+2\partial_i\left(\partial^k\zeta_n\partial^i\zeta_n\right)\partial^{-2}\partial_l\left(\partial_k\zeta_n\partial^l\zeta_n\right)
+\frac{1}{2}\left(\partial^{-2}\partial_j\partial_k\left(\partial^j\zeta_n\partial^k\zeta_n\right)\right)^2
\Big]\nonumber
\\&&-\frac{5\epsilon}{8a^3H^4}\Big[\partial_l\zeta_n\partial^l\zeta_n\partial^k\partial^j\left(\partial_k\zeta_n\partial_j\zeta_n\right)
-
\frac{1}{2}\partial_j\zeta_n\partial^j\zeta_n\partial^2\left(\partial_k\zeta_n\partial^k\zeta_n\right)
\nonumber
\\&&\;\;\;\;\;\;\;\;\;\;\;\;\;\;\;\;\;\;-\frac{1}{2}\partial^k\partial^j\left(\partial_k\zeta_n\partial_j\zeta_n\right)\partial^{-2}\partial^m\partial^l\left(\partial_m\zeta_n\partial_l\zeta_n\right)
\Big]\Big\}.\label{Sparts}
\end{eqnarray}
For us to be able to obtain the previous result it is crucial to
include the contributions from the tensor actions (\ref{GG}) and
(\ref{GSS}), otherwise the trispectrum calculated using the
$\mathcal{O}(\epsilon^0)$ of (\ref{SSSS}) does not vanish, giving
the wrong leading order result. Neglecting tensor perturbations
(sourced by the scalars) for the calculation of the trispectrum is
not consistent and leads to wrong results. This is one of the
results of our work. The contribution from $\tilde{\gamma}_{ij}$
that comes through $\gamma_{ij}$ in (\ref{GG}) and (\ref{GSS})
will result in terms that are already slow-roll suppressed and no
further integrations by parts are need on these terms (see Eq.
(\ref{tildegammaeq}) of next subsection). The final action for
$\zeta_n$ is then
\begin{equation}
{S_4}_{\zeta_n}=S_{\zeta_n}^{(\epsilon)}+S_{parts},\label{zetanactionsup}
\end{equation}
where $S_{\zeta_n}^{(\epsilon)}$ denotes the terms of
(\ref{zetanaction}) that are suppressed by at least one slow-roll
parameter. This final action is slow-roll suppressed as expected
and no slow-roll approximation was made so it is also exact.

\subsection{\label{subsec:PerturbationsUniformeR}Non-linear
perturbations in the uniform curvature gauge}

In order to calculate the intrinsic four point correlation
function of the field perturbation we need to compute the action
of fourth order in the perturbations. In this subsection we will
obtain the fourth order action in the uniform curvature gauge. In
this gauge, the scalar degree of freedom is the inflaton field
perturbation $\delta\phi(x^\mu)$. There are several works in the
literature where the authors also calculate the trispectrum. In
\cite{Seery:2006vu}, Seery \emph{et al.} calculate the trispectrum
of a multi-field inflation model, however the result is only valid
for fields with standard kinetic energies, i.e.
$P(X_1,\ldots,X_n,\phi_1,\ldots,\phi_n)=X_1+\cdots+X_n-V$, where
$X_n$ is the kinetic energy of $\phi_n$ and $V$ is the potential.
In this paper we will generalize their result for an arbitrary
function $P(X,\phi)$ but for single field only. Recently, Huang
and Shiu have obtained the fourth order action for the model under
consideration (\ref{action}). However the result was obtained by
only perturbing the field lagrangian. This procedure gives the
right result as long as we are interested in the leading order
contribution in the small speed of sound limit and in the
slow-roll approximation, which was their case of interest. In the
present section we will compute the fourth order action that is
valid to all orders in slow roll and in the sound of speed
expansion. To do that we have to perturb the full action
(\ref{action}) up to fourth order in the field perturbations. The
procedure to obtain the fourth order action in this gauge is
similar to the one used in subsection
\ref{subsec:PerturbationsComoving}.

In this gauge, the inflaton perturbation does not vanish and the
3D metric takes the form
\begin{eqnarray}
&&\phi(x,t)=\phi_0+\delta\phi(x,t),\nonumber\\
&&h_{ij}=a^2\hat{h}_{ij}, \quad
\hat{h}_{ij}=\delta_{ij}+\tilde{\gamma}_{ij}+\frac{1}{2}\tilde{\gamma}_{ik}\tilde{\gamma}_j^k+\cdots\label{deltaphigauge}
\end{eqnarray}
where $\mbox{det} \hat{h}=1$ and $\tilde{\gamma}_{ij}$ is a tensor
perturbation that we assume to be a second order quantity, i.e.,
$\tilde{\gamma}_{ij}=\mathcal{O}(\delta\phi^2)$. It obeys the
traceless and transverse conditions
$\tilde{\gamma}_i^i=\partial^i\tilde{\gamma}_{ij}=0$ (indices are
raised with $\delta_{ij}$). In the literature the second order
tensor perturbations are often neglected, however based on our
results it should be taken into account. We can always use the
gauge freedom at second order to eliminate the trace and the
vector perturbations of $h_{ij}$. The presence of
$\tilde{\gamma}_{ij}$ makes the three dimensional hypersurfaces
non-flat so using the name uniform curvature gauge might be
misleading. We will continue to use that name because in the
literature that is the name given to the gauge where
$\delta\phi(x,t)=0$.

We expand $N$ and $N^i$ in powers of the perturbation
$\delta\phi(x,t)$
\begin{eqnarray}
N=1+\alpha_1+\alpha_2+\cdots,\\
\tilde{N_i}=\tilde{N_i}^{(1)}+\tilde{N_i}^{(2)}+\cdots,\\
\psi=\psi_1+\psi_2+\cdots,
\end{eqnarray}
where $\alpha_n$, $\tilde{N_i}^{(n)}$ and $\psi_n$ are of order
$\delta\phi^n$ and $\phi_0(t)$ is the background value of the
field. At first order in $\delta\phi$, a particular solution for
equations (\ref{LMphi}) is \cite{Maldacena:2002vr,Seery:2006vu}:
\begin{equation}
\alpha_1=\frac{1}{2H}\dot{\phi_0}\delta\phi P_{,X}, \quad
\tilde{N_i}^{(1)}=0, \quad
\partial^2\psi_1=\frac{a^2\epsilon}{c_s^2}\frac{d}{dt}\left(-\frac{H}{\dot{\phi}}\delta\phi\right)
. \label{N1orderphi}
\end{equation}
At second order, the constraint equation for the lapse function
gives
\begin{eqnarray}
\frac{4H}{a^2}\partial^2\psi_2&=&\frac{1}{a^4}\left(\partial^2\psi_1\partial^2\psi_1-\partial_i\partial_j\psi_1\partial^i\partial^j\psi_1\right)
+6H^2\left(3\alpha_1^2-2\alpha_2\right)\nonumber\\\nonumber
&&+\frac{8H\alpha_1}{a^2}\partial^2\psi_1-\left(3\dot{\phi_0}^2\alpha_1^2-2\dot{\phi_0}^2\alpha_2+\dot{\delta\phi}^2-4\dot{\phi_0}\dot{\delta\phi}\alpha_1\right)P_{,X}
-\frac{1}{a^2}\left(\partial_i\delta\phi\partial^i\delta\phi-2\dot{\phi_0}\partial_i\delta\phi\partial^i\psi_1\right)P_{,X}\\\nonumber
&&+\delta\phi^2P_{,\phi\phi}+2\dot{\phi_0}\delta\phi\left(\dot{\phi_0}\alpha_1-\dot{\delta\phi}\right)P_{,X\phi}\\\nonumber
&&-\frac{\dot{\phi_0}^2}{a^2}\left(-10a^2\dot{\phi_0}\dot{\delta\phi}\alpha_1-2\dot{\phi_0}\partial_i\delta\phi\partial^i\psi_1
+6a^2\dot{\phi_0}^2\alpha_1^2-2\dot{\phi_0}^2a^2\alpha_2+4a^2\dot{\delta\phi}^2-\partial_i\delta\phi\partial^i\delta\phi\right)P_{,XX}\\
&&-\dot{\phi_0}^4\left(\dot{\phi_0}^2\alpha_1^2+\dot{\delta\phi}^2-2\dot{\phi_0}\dot{\delta\phi}\alpha_1\right)P_{,XXX}
+2\dot{\phi_0}^3\delta\phi\left(\dot{\phi_0}\alpha_1-\dot{\delta\phi}\right)P_{,XX\phi}-\dot{\phi_0}^2\delta\phi^2P_{,X\phi\phi},
\end{eqnarray}
and the equation for the shift vector gives
\begin{eqnarray}
2H\partial_i\alpha_2-\frac{1}{2}a^{-2}\partial^2\tilde{N_i}^{\!\!(2)}&=&
4\alpha_1H\partial_i\alpha_1+a^{-2}\left(\partial_i\alpha_1\partial^2\psi_1-\partial_k\alpha_1\partial_i\partial^k\psi_1\right)
+\partial_i\delta\phi\left(\dot{\delta\phi}-\dot{\phi_0}\alpha_1\right)P_{,X}\nonumber\\
&&+\partial_i\delta\phi\dot{\phi_0}^2\left(\dot{\delta\phi}-\dot{\phi_0}\alpha_1\right)P_{,XX}
+\partial_i\delta\phi\dot{\phi_0}\delta\phi P_{,X\phi},
\label{Niconstraint2orderphi}
\end{eqnarray}
where $P_{,\phi}$ means derivative of $P$ with respect to $\phi$.

Due to the fact that $\tilde{N^i}$ is divergenceless and that any
vector can be separated into a incompressible and irrotational
part one can separate the contributions from $\alpha_2$ and
$\tilde{N_i}^{\!\!(2)}$ in the previous equation. The irrotational
part of Eq. (\ref{Niconstraint2orderphi}) gives
\begin{equation}
2H\alpha_2=\partial^{-2}\partial^iF_i, \label{alpha2phi}
\end{equation}
and the incompressible part gives
\begin{equation}
\frac{1}{2a^2}\tilde{N_i}^{\!\!(2)}=-\partial^{-2}F_i+
\partial^{-4}\partial_i\partial^kF_k,
\label{Ni2phi}
\end{equation}
where $F_i$ is defined as the right-hand side of equation
(\ref{Niconstraint2orderphi}).

The scalar fourth order action, where no slow-roll approximation
has been made, is
\begin{equation}
S_4=S_A+S_B,\label{SSSSphi}
\end{equation}
where
\begin{eqnarray}
S_A&=&\int dtd^3x\Big[ -\frac {a\delta\phi}{2} \big[
a^{2}\alpha_1\dot{\delta\phi}^2+2\partial_i\delta\phi
\partial^i\psi_1\left(\dot{\delta\phi}-\alpha_1\dot{\phi}_0\right)+2\dot{\phi}_0\partial^i\delta\phi\left(\tilde{N_i}^{\!\!(2)}
+\partial_i\psi_2\right)+a^2\dot{\phi}_0^2\alpha_1^3+\alpha_1(\partial\delta\phi)^2
\big] P_{,X\phi} \nonumber
\\
\nonumber && -\frac{1}{8a} \big[
4\dot{\phi}_0^4a^4\alpha_2\left(\alpha_2-2\alpha_1^2\right)+3\dot{\phi}_0^4a^4\alpha_1^4-18\dot{\phi}_0^2a^4\alpha_1^2\dot{\delta\phi}^2+8\dot{\phi}_0^2a^2\partial^i\delta\phi
\left(\tilde{N_i}^{\!\!(2)}+
\partial_i\psi_2\right)\left(\dot{\delta\phi}-\dot{\phi}_0\alpha_1\right)\\\nonumber
&&-4\dot{\phi}_0\partial_i\delta\phi\partial^i\psi_1\left((\partial\delta\phi)^2-3\dot{\phi}_0^2a^2\alpha_1^2+8\dot{\phi}_0a^2\alpha_1\dot{\delta\phi}-3a^2\dot{\delta\phi}^2\right)
+12\dot{\phi}_0a^4
\alpha_1\dot{\delta\phi}^3+4\dot{\phi}_0^3a^4\alpha_1^3
\dot{\delta\phi}
\\\nonumber
&&-4\,\dot{\phi}_0^2\partial_i\delta\phi\partial^i\psi_1\partial_k\delta\phi
\partial^k\psi_1-2 \dot{\phi}_0 ^2a^2 \alpha_1^2
(\partial\delta\phi)^2 -4 \alpha_1\dot{\phi}_0\dot{\delta\phi}a^2
(\partial\delta\phi)^2
-a^4\left(\dot{\delta\phi}^2-a^{-2}(\partial\delta\phi)^2\right)^2\big]P_{,XX}
\\
\nonumber &&
+\frac{\dot{\phi}_0^2a}{12}\big[6\dot{\phi}_0\partial_i\delta\phi
\partial^i\psi_1\left(2\dot{\phi}_0\alpha_1\dot{\delta\phi}-\dot{\delta\phi}^2-\dot{\phi}_0^2\alpha_1^2\right)
+3(\partial\delta\phi)^2\left(2\dot{\phi}_0\alpha_1\dot{\delta\phi}-\dot{\phi}_0^2\alpha_1^2\right)
-16\dot{\phi}_0a^2\alpha_1\dot{\delta\phi}^3
\\\nonumber &&
+24\dot{\phi}_0^2a^2\alpha_1^2\dot{\delta\phi}^2-12\dot{\phi}_0^3a^2\alpha_1^3\dot{\delta\phi}+\dot{\phi}_0^4a^2\alpha_1^4
+3a^2\dot{\delta\phi}^2\left(\dot{\delta\phi}^2-a^{-2}(\partial\delta\phi)^2\right)\big]
P_{,XXX}
\\
\nonumber && +\frac{\dot{\phi}_0a\delta\phi}{2}\big[
a^2\dot{\delta\phi}^3+\left((\partial\delta\phi)^2+2\dot{\phi}_0\partial_i\delta\phi\partial^i\psi_1\right)\left(\dot{\phi}_0\alpha_1-\dot{\delta\phi}\right)
+3\dot{\phi}_0^2a^2\alpha_1^2\dot{\delta\phi}-4\dot{\phi}_0a^2
\alpha_1 \dot{\delta\phi}^2 \big] P_{,XX\phi}
\\
\nonumber && -\frac{a\delta\phi^2}{8}\big[
2(\partial\delta\phi)^2-2a^2\dot{\delta\phi}^2+4a^2\alpha_1\dot{\phi}_0\dot{\delta\phi}+2a^2\dot{\phi}_0^2\alpha_1^2+4
\dot{\phi}_0\partial_i\delta\phi\partial^i\psi_1 \big]
P_{,X\phi\phi}
\\
\nonumber && +\frac{1}{24}\dot{\phi}_0^4a^3 \big[
6\dot{\phi}_0^2\alpha_1^2\dot{\delta\phi}^2-4\dot{\phi}_0\alpha_1
\dot{\delta\phi}^3-4\dot{\phi}_0^3\alpha_1^3\dot{\delta\phi}+\dot{\phi}_0^4
\alpha_1^4+\dot{\delta\phi}^4 \big]
 P_{,XXXX}
\\
\nonumber &&-\frac{1}{6}\dot{\phi}_0^3\delta\phi a^3 \left(
-\dot{\delta\phi}^3+3\dot{\phi}_0\alpha_1\dot{\delta\phi}^2+\dot{\phi}_0^3\alpha_1^3-3\dot{\phi}_0^2\alpha_1^2\dot{\delta\phi}
\right) P_{,XXX\phi}
\\
 &&+\frac{1}{4}\dot{\phi}_0^2 \delta\phi^2a^3
 \left( -2\alpha_1\dot{\phi}_0\dot{\delta\phi}+\dot{\delta\phi}^2+\dot{\phi}_0^2\alpha_1^2
\right)P_{,XX\phi\phi} -\frac{1}{6}\dot{\phi}_0\delta\phi^3a^3
\left(\alpha_1\dot{\phi}_0-\dot{\delta\phi}\right)P_{,X\phi\phi\phi}
\Big], \label{SSSSphiA}
\end{eqnarray}
\begin{eqnarray}
S_B&=&\int dtd^3x\Big[\alpha_1^3a^3\delta\phi P_{,\phi}
+\frac{1}{2}a^3\alpha_1^2\delta\phi^2P_{,\phi\phi}+\frac{1}{6}a^3\alpha_1\delta\phi^3P_{,\phi\phi\phi}+\frac{1}{24}a^3\delta\phi^4P_{,4\phi}
\nonumber\\\nonumber
&&-\frac{1}{2a}\big[-(\partial_i\delta\phi\partial^i\psi_1)^2-2\alpha_1\dot{\delta\phi}a^2\partial_i\delta\phi\partial^i\psi_1
+2a^2\partial_i\delta\phi\left(\tilde{N_i}^{\!\!(2)}+\partial_i\psi_2\right)\left(\dot{\delta\phi}-\alpha_1\dot{\phi}_0\right)
\\\nonumber &&+\alpha_1^2a^2(\partial\delta\phi)^2+
\dot{\phi}_0^2a^4\alpha_2\left(\alpha_2-2\alpha_1^2\right)\big]P_{,X}
\\
&&+\frac{1}{4a}\big[2\partial^i\tilde{N^j}^{\!\!(2)}\partial_{(i}\tilde{N_{j)}}^{\!\!(2)}-4\alpha_1\partial_i\partial_k\psi_1\left(\partial^i\partial^k\psi_2
+\partial^i\tilde{N^k}^{\!\!(2)}\right)+12a^4H^2\alpha_2\left(\alpha_2-2\alpha_1^2\right)+4\alpha_1\partial^2\psi_1\partial^2\psi_2\big]
\Big]. \label{SSSSphiB}
\end{eqnarray}
The previous actions should be supplemented with the pure tensor
terms and the tensor-scalar coupling terms:
\begin{equation}
S=\frac{1}{8}\int
dtd^3x\left[a^3\dot{\tilde{\gamma}}_{ij}\dot{\tilde{\gamma}}^{ij}-a\partial_k\tilde{\gamma}_{ij}\partial^k\tilde{\gamma}^{ij}\right],
\label{GGphi}
\end{equation}
\begin{equation}
S=\int
dtd^3x\left[aP_{,X}\tilde{\gamma}^{ij}\partial_j\delta\phi\left(\frac{1}{2}\partial_i\delta\phi+\dot{\phi_0}\partial_i\psi_1\right)\right].
\label{GSSphi}
\end{equation}

This constitutes the main result of this subsection. It is a good
check for our calculation to see that the previous action
(\ref{SSSSphi}) reduces in some particular cases to previously
know results present in the literature.

For example, if we restrict our model to the standard inflation
case, i.e., $P(X,\phi)=X-V(\phi)$, where $V(\phi)$  is the
inflaton potential, then all the terms in the scalar action
(\ref{SSSSphiA}) vanish and the only contribution to the fourth
order action comes from (\ref{SSSSphiB}). These terms exactly
reproduce the result of Seery \emph{et al.} \cite{Seery:2006vu},
their equation (36), restricted to single field. However, in the
total fourth order action there are also the tensor contributions
(\ref{GGphi}) and (\ref{GSSphi}). In general, to proceed one has
to calculate the equation of motion for the second order tensor
perturbations $\tilde{\gamma}_{ij}$ from eqs. (\ref{GGphi}),
(\ref{GSSphi}) to get
\begin{equation}
\tilde{\gamma}_{ij}''+2\frac{a'}{a}\tilde{\gamma}_{ij}'-\partial^2\tilde{\gamma}_{ij}=
\left(2P_{,X}\partial_j\delta\phi\partial_i\delta\phi+4P_{,X}\dot{\phi_0}\partial_j\delta\phi\partial_i\psi_1\right)^{TT},
\label{tildegammaeq}
\end{equation}
where TT means the transverse and traceless parts of the
expression inside the parenthesis (see Appendix \ref{TTpart} for
details of how to extract the TT parts of a tensor) and then solve
this equation to obtain $\tilde{\gamma}_{ij}$ as a function of
$\delta\phi$. One can immediately see that at second order the
scalars will source the tensor perturbation equation as it was
previously shown by others \cite{Ananda:2006af,Osano:2006ew}. At
this order in perturbation theory, equation (\ref{tildegammaeq})
should also have a source term quadratic in the first order tensor
perturbations, ${{}^{(1)}\tilde{\gamma}_{ij}}_{GW}$. We neglect
these terms because we expect that any correlation function where
${{}^{(2)}\tilde{\gamma}_{ij}}_{GW}$ enters, which is sourced by
the first order tensor modes squared, must be smaller than a
correlation function with only scalars, see Ref.
\cite{Maldacena:2002vr} for an example. In Fourier space, the
source term of (\ref{tildegammaeq}) is suppressed by $k^2$, where
$k$ is the wave number. Once we have the solution of
$\tilde{\gamma}_{ij}$ in terms of $\delta\phi$ we can plug back
the result in (\ref{GGphi}) and (\ref{GSSphi}) to get the total
fourth order scalar action.

\section{\label{sec:trispectrumformalism}The general formalism to calculate the trispectrum}
\subsection{\label{subsec:trispectrumzetan}The trispectrum of $\zeta_n$}

Now we shall give the basic equations needed to calculate the
trispectrum \cite{Maldacena:2002vr,Weinberg:2005vy}. First we need
to solve the second order equation of motion for $\zeta_n$
(obtained from (\ref{2action})). Defining new variables
\begin{equation}
v_k=zu_k, \quad z=\frac{a\sqrt{2\epsilon}}{c_s},
\end{equation}
where the Fourier mode function $u_k$ is given by
\begin{equation}
u_k=\int d^3x\zeta_n(t,\B{x})e^{-i\B{k}\cdot\B{x}},
\end{equation}
the equation of motion for $\zeta_n$ is
\begin{equation}
v_k''+c_s^2k^2v_k-\frac{z''}{z}v_k=0, \label{Mukhanoveq}
\end{equation}
where prime denotes derivative with respect to conformal time
$\tau$. This is also known as the Mukhanov equation. The previous
equation can be solved, at leading order in slow roll and if the
rate of change of the sound speed is small \cite{Chen:2006nt}, to
give
\begin{equation}
u_k\equiv u(\tau,\B{k})=\frac{iH}{\sqrt{4\epsilon
c_sk^3}}\left(1+ikc_s\tau\right)e^{-ikc_s\tau}.\label{modefc}
\end{equation}
We do not need to impose any constraints in the sound speed and it
can be arbitrary. Only its rate of change is assumed to be small.
The next-to-leading order corrections to the previous solutions
are also known and can be found in \cite{Chen:2006nt}. In the
general case, we would have to solve Eq. (\ref{Mukhanoveq})
without assuming slow roll. This can be done numerically.

In order to calculate the $\zeta_n$ correlators we follow the
standard procedure in quantum field theory. The curvature
perturbation is promoted to an operator that can be expanded in
terms of creation and annihilation operator as
\begin{equation}
\zeta_n(\tau,\B{k})=u(\tau,\B{k})a(\B{k})+u^*(\tau,-\B{k})a^\dag(-\B{k}).
\end{equation}
The standard commutation relation applies
\begin{equation}
\left[a(\B{k_1}),a^\dag(\B{k_2})\right]=(2\pi)^3\delta^{(3)}(\B{k_1}-\B{k_2}).
\end{equation}
The vacuum expectation value of the four point operator in the
interaction picture (at first order) is
\cite{Maldacena:2002vr,Weinberg:2005vy}
\begin{equation}
\langle\Omega|\zeta_n(t,\B{k_1})\zeta_n(t,\B{k_2})\zeta_n(t,\B{k_3})\zeta_n(t,\B{k_4})|\Omega\rangle=
-i\int_{t_0}^td\tilde{t}\langle 0|
\left[\zeta_n(t,\B{k_1})\zeta_n(t,\B{k_2})\zeta_n(t,\B{k_3})\zeta_n(t,\B{k_4}),H_I(\tilde{t})\right]|0\rangle,
\label{vev}
\end{equation}
where $t_0$ is some early time during inflation when the inflaton
vacuum fluctuation is deep inside the horizon, $t$ is some time
after horizon exit. $|\Omega\rangle$ is the interacting vacuum
which is different from the free theory vacuum $|0\rangle$. If one
uses conformal time, it's a good approximation to perform the
integration from $-\infty$ to $0$ because $\tau\approx-(aH)^{-1}$.
$H_I$ denotes the interaction hamiltonian and it is given by
$H_I=\pi\dot\zeta_n-L$, where $\pi$ is defined as $\pi=\frac{\partial L}{\partial \dot\zeta_n}$ and $L$ is the lagrangian. In this work, we will only calculate the contribution for the four point function
that comes from a part of the interaction Hamiltonian determined by the 4th order Lagrangian
$H_I=-L_4$, where $L_4$ is the total lagrangian obtained from the
action (\ref{zetanactionsup}). We should point out that the other terms that we do not consider
here in the fourth order interaction hamiltonian are indeed important to obtain the full leading
order result (see equation (\ref{trizetan})) as was recently shown by \cite{Huang:2006eh}.

Of course in the end we are
interested in the four point correlation function of $\zeta$ and
not of $\zeta_n$. At leading order in slow roll these two
correlation functions are equal but they will differ at
next-to-leading order.

\subsection{\label{subsec:trispectrumzeta}The trispectrum of $\zeta$}

In this subsection we calculate the relation between the
trispectrum of $\zeta$, on large scales, and the trispectrum  of
$\zeta_n$ calculated using the formalism of the previous
subsection. This relation also involves lower-order correlation
functions of $\zeta_n$ present in the literature. The variables
$\zeta$ and $\zeta_n$ are related up to third order by
\begin{equation}
\zeta(\B{x})=\zeta_n(\B{x})+F_2(\zeta_n(\B{x}))+F_3(\zeta_n(\B{x})),\label{gte}
\end{equation}
where $F_2(\zeta_n)=\mathcal{O}(\zeta_n^2)$,
$F_3(\zeta_n)=\mathcal{O}(\zeta_n^3)$ are the terms coming from
the second and third order gauge transformations respectively.
$F_2$ can be found in Appendix \ref{GT}, it is
\begin{equation}
F_2(\zeta_n)=\left(\frac{\epsilon}{2}+\frac{\ddot{\phi}_0}{2H\dot{\phi}_0}\right)\zeta_n^2
+\frac{1}{H}\zeta_n\dot{\zeta}_n+\beta,
\end{equation}
where $\beta$ is given in Eq. (\ref{beta}). In the large scale
limit (super-horizon scales), we can ignore $\beta$ as it contains
gradient terms. $F_3$ was calculated in \cite{Jarnhus:2007ia} and
reads
\begin{equation}
F_3(\zeta_n)=\left(\frac{\dddot{\phi}_0}{3H^2\dot{\phi}_0}+\frac{\epsilon\ddot{\phi}_0}{H\dot{\phi}_0}+
\frac{\epsilon^2}{3}+\frac{\epsilon\eta}{3}\right)\zeta_n^3+\left(\frac{3\ddot{\phi}_0}{2H\dot{\phi}_0}+
2\epsilon\right)\frac{\dot{\zeta}_n\zeta_n^2}{H}+\frac{\zeta_n\dot{\zeta}_n^2}{H^2}
+ \frac{\ddot{\zeta}_n\zeta_n^2}{2H^2} + f_a(\zeta_n) +
f_b(\zeta_n,\tilde{\gamma}_{ij}),
\end{equation}
where $f_a$ denotes terms that contain gradients (it can be found
in \cite{Jarnhus:2007ia}). $f_b$ is the part of the third order
gauge transformations that contains $\tilde{\gamma}_{ij}$. The
explicit form of $f_b$ is to the best of our knowledge still
unknown. To find out the explicit dependence of these terms on
$\zeta_n$ one would have to solve the equations of motion for
$\tilde{\gamma}_{ij}$, equation (\ref{tildegammaeq}). We do not do
this in this work. We believe that these terms will vanish in the
large scale limit and therefore do not contribute to our
calculation.

A field redefinition like
$\zeta=\zeta_n+a_1\zeta_n^{(a)}\zeta_n^{(b)}+a_2\zeta_n^{(c)}\zeta_n^{(d)}$,
where $\zeta_n^{(a,b,c,d)}$ denote one of $\zeta_n, \dot{\zeta}_n,
\ddot{\zeta}_n$, gives after using Wick's theorem a relation
between both trispectrum like
\begin{equation}
\langle\zeta(\B{x}_1)\zeta(\B{x}_2)\zeta(\B{x}_3)\zeta(\B{x}_4)\rangle_c=\{T\}+\{PB\}+\{PPP\}+\mathcal{O}(P^\zeta_k)^4,
\end{equation}
where
\begin{equation}
\{T\}=\langle\zeta_n(\B{x}_1)\zeta_n(\B{x}_2)\zeta_n(\B{x}_3)\zeta_n(\B{x}_4)\rangle,
\end{equation}
\begin{eqnarray}
\{PB\}&=&a_1\Big[\langle\zeta_n^{(a)}(\B{x}_1)\zeta_n(\B{x}_2)\rangle\langle\zeta_n^{(b)}(\B{x}_1)\zeta_n(\B{x}_3)\zeta_n(\B{x}_4)\rangle
+\langle\zeta_n^{(a)}(\B{x}_1)\zeta_n(\B{x}_3)\rangle\langle\zeta_n^{(b)}(\B{x}_1)\zeta_n(\B{x}_2)\zeta_n(\B{x}_4)\rangle
\nonumber\\&&
\quad+\langle\zeta_n^{(a)}(\B{x}_1)\zeta_n(\B{x}_4)\rangle\langle\zeta_n^{(b)}(\B{x}_1)\zeta_n(\B{x}_2)\zeta_n(\B{x}_3)\rangle
+\langle\zeta_n^{(b)}(\B{x}_1)\zeta_n(\B{x}_2)\rangle\langle\zeta_n^{(a)}(\B{x}_1)\zeta_n(\B{x}_3)\zeta_n(\B{x}_4)\rangle
\nonumber\\&&
\quad+\langle\zeta_n^{(b)}(\B{x}_1)\zeta_n(\B{x}_3)\rangle\langle\zeta_n^{(a)}(\B{x}_1)\zeta_n(\B{x}_2)\zeta_n(\B{x}_4)\rangle
+\langle\zeta_n^{(b)}(\B{x}_1)\zeta_n(\B{x}_4)\rangle\langle\zeta_n^{(a)}(\B{x}_1)\zeta_n(\B{x}_2)\zeta_n(\B{x}_3)\rangle
\nonumber\\&& \quad+3\;perm. \Big]
\nonumber\\&+&a_2\Big[(a\rightarrow c,b\rightarrow
d)+3\;perm.\Big],
\end{eqnarray}
\begin{eqnarray}
\{PPP\}&=&a_1^2\Big[\langle\zeta_n^{(a)}(\B{x}_1)\zeta_n^{(a)}(\B{x}_2)\rangle\Big(\langle\zeta_n^{(b)}(\B{x}_1)\zeta_n(\B{x}_3)\rangle\langle\zeta_n^{(b)}(\B{x}_2)\zeta_n(\B{x}_4)\rangle
+\langle\zeta_n^{(b)}(\B{x}_1)\zeta_n(\B{x}_4)\rangle\langle\zeta_n^{(b)}(\B{x}_2)\zeta_n(\B{x}_3)\rangle\Big)
\nonumber\\&& \quad+
\langle\zeta_n^{(a)}(\B{x}_1)\zeta_n^{(b)}(\B{x}_2)\rangle\Big(\langle\zeta_n^{(b)}(\B{x}_1)\zeta_n(\B{x}_3)\rangle\langle\zeta_n^{(a)}(\B{x}_2)\zeta_n(\B{x}_4)\rangle
+\langle\zeta_n^{(b)}(\B{x}_1)\zeta_n(\B{x}_4)\rangle\langle\zeta_n^{(a)}(\B{x}_2)\zeta_n(\B{x}_3)\rangle\Big)
\nonumber\\&& \quad+
\langle\zeta_n^{(b)}(\B{x}_1)\zeta_n^{(a)}(\B{x}_2)\rangle\Big(\langle\zeta_n^{(a)}(\B{x}_1)\zeta_n(\B{x}_3)\rangle\langle\zeta_n^{(b)}(\B{x}_2)\zeta_n(\B{x}_4)\rangle
+\langle\zeta_n^{(a)}(\B{x}_1)\zeta_n(\B{x}_4)\rangle\langle\zeta_n^{(b)}(\B{x}_2)\zeta_n(\B{x}_3)\rangle\Big)
\nonumber\\&& \quad+
\langle\zeta_n^{(b)}(\B{x}_1)\zeta_n^{(b)}(\B{x}_2)\rangle\Big(\langle\zeta_n^{(a)}(\B{x}_1)\zeta_n(\B{x}_3)\rangle\langle\zeta_n^{(a)}(\B{x}_2)\zeta_n(\B{x}_4)\rangle
+\langle\zeta_n^{(a)}(\B{x}_1)\zeta_n(\B{x}_4)\rangle\langle\zeta_n^{(a)}(\B{x}_2)\zeta_n(\B{x}_3)\rangle\Big)
\nonumber\\&& \quad+ 5\;perm. \Big]
\nonumber\\&+&a_2^2\Big[(a\rightarrow c,b\rightarrow
d)+5\;perm.\Big]
\nonumber\\&+&2a_1a_2\Big[\langle\zeta_n^{(a)}(\B{x}_1)\zeta_n^{(c)}(\B{x}_2)\rangle\Big(\langle\zeta_n^{(b)}(\B{x}_1)\zeta_n(\B{x}_3)\rangle\langle\zeta_n^{(d)}(\B{x}_2)\zeta_n(\B{x}_4)\rangle
+\langle\zeta_n^{(b)}(\B{x}_1)\zeta_n(\B{x}_4)\rangle\langle\zeta_n^{(d)}(\B{x}_2)\zeta_n(\B{x}_3)\rangle\Big)
\nonumber\\&& \quad+
\langle\zeta_n^{(a)}(\B{x}_1)\zeta_n^{(d)}(\B{x}_2)\rangle\Big(\langle\zeta_n^{(b)}(\B{x}_1)\zeta_n(\B{x}_3)\rangle\langle\zeta_n^{(c)}(\B{x}_2)\zeta_n(\B{x}_4)\rangle
+\langle\zeta_n^{(b)}(\B{x}_1)\zeta_n(\B{x}_4)\rangle\langle\zeta_n^{(c)}(\B{x}_2)\zeta_n(\B{x}_3)\rangle\Big)
\nonumber\\&& \quad+
\langle\zeta_n^{(b)}(\B{x}_1)\zeta_n^{(c)}(\B{x}_2)\rangle\Big(\langle\zeta_n^{(a)}(\B{x}_1)\zeta_n(\B{x}_3)\rangle\langle\zeta_n^{(d)}(\B{x}_2)\zeta_n(\B{x}_4)\rangle
+\langle\zeta_n^{(a)}(\B{x}_1)\zeta_n(\B{x}_4)\rangle\langle\zeta_n^{(d)}(\B{x}_2)\zeta_n(\B{x}_3)\rangle\Big)
\nonumber\\&& \quad+
\langle\zeta_n^{(b)}(\B{x}_1)\zeta_n^{(d)}(\B{x}_2)\rangle\Big(\langle\zeta_n^{(a)}(\B{x}_1)\zeta_n(\B{x}_3)\rangle\langle\zeta_n^{(c)}(\B{x}_2)\zeta_n(\B{x}_4)\rangle
+\langle\zeta_n^{(a)}(\B{x}_1)\zeta_n(\B{x}_4)\rangle\langle\zeta_n^{(c)}(\B{x}_2)\zeta_n(\B{x}_3)\rangle\Big)
\nonumber\\&& \quad+ 5\;perm. \Big],
\end{eqnarray}
where ``perm" means the other permutations of the preceding terms
and $\mathcal{O}(P^\zeta_k)^4$ denotes terms that are suppressed
by successive powers of the power spectrum. $(a\rightarrow
c,b\rightarrow d)$ means terms equal to the immediately preceding
terms with $a,b$ replaced by $c,d$ respectively.

If the field redefinition contains third order pieces like
$\zeta=\zeta_n+b_1\zeta_n^{(a)}\zeta_n^{(b)}\zeta_n^{(c)}$, they
contribute with additional terms as
\begin{eqnarray}
\langle\zeta(\B{x}_1)\zeta(\B{x}_2)\zeta(\B{x}_3)\zeta(\B{x}_4)\rangle_c&=&\langle\zeta_n(\B{x}_1)\zeta_n(\B{x}_2)\zeta_n(\B{x}_3)\zeta_n(\B{x}_4)\rangle
\nonumber\\&+&b_1\Big[\langle\zeta_n^{(a)}(\B{x}_1)\zeta_n(\B{x}_2)\rangle\Big(\langle\zeta_n^{(b)}(\B{x}_1)\zeta_n(\B{x}_3)\rangle\langle\zeta_n^{(c)}(\B{x}_1)\zeta_n(\B{x}_4)\rangle
+\langle\zeta_n^{(c)}(\B{x}_1)\zeta_n(\B{x}_3)\rangle\langle\zeta_n^{(b)}(\B{x}_1)\zeta_n(\B{x}_4)\rangle\Big)
\nonumber\\&& \quad+
\langle\zeta_n^{(b)}(\B{x}_1)\zeta_n(\B{x}_2)\rangle\Big(\langle\zeta_n^{(a)}(\B{x}_1)\zeta_n(\B{x}_3)\rangle\langle\zeta_n^{(c)}(\B{x}_1)\zeta_n(\B{x}_4)\rangle
+\langle\zeta_n^{(c)}(\B{x}_1)\zeta_n(\B{x}_3)\rangle\langle\zeta_n^{(a)}(\B{x}_1)\zeta_n(\B{x}_4)\rangle\Big)
\nonumber\\&& \quad+
\langle\zeta_n^{(c)}(\B{x}_1)\zeta_n(\B{x}_2)\rangle\Big(\langle\zeta_n^{(a)}(\B{x}_1)\zeta_n(\B{x}_3)\rangle\langle\zeta_n^{(b)}(\B{x}_1)\zeta_n(\B{x}_4)\rangle
+\langle\zeta_n^{(b)}(\B{x}_1)\zeta_n(\B{x}_3)\rangle\langle\zeta_n^{(a)}(\B{x}_1)\zeta_n(\B{x}_4)\rangle\Big)
\nonumber\\&& \quad+3\;perm. \Big] + \mathcal{O}(P^\zeta_k)^4.
\end{eqnarray}
To the best of our knowledge the expectation values involving
operators containing derivatives of $\zeta_n$ have not yet been
calculated in the literature. However, once the mode function
equation (\ref{Mukhanoveq}) is solved, one has all the ingredients
needed to calculate these expectation values, including the
interaction hamiltonian.

\section{\label{sec:leadingtrispectrum}Calculation of the leading
order trispectrum}

In this section, we will use the formalism of the previous section
and the fourth order exact interaction hamiltonian of subsection
\ref{subsec:PerturbationsComoving} to calculate the leading order
trispectrum, under the assumption that the ``slow-roll" parameters
(\ref{epsilon}-\ref{s}) are always small until the end of
inflation.

\subsection{\label{subsec:leadingtrispectrumzetan}The leading order
trispectrum of $\zeta_n$}

To calculate the leading order trispectrum of $\zeta_n$ in slow
roll, we need to evaluate Eq. (\ref{vev}) where $H_I$ is read from
the order $\epsilon$ terms of the action (\ref{zetanactionsup}).
The interaction hamiltonian we get contains terms with
$\tilde{\gamma}_{ij}$. Fortunately it can be shown that to compute
the leading order trispectrum we don't need to know the solution
for $\tilde{\gamma}_{ij}$ and the knowledge of its equation of
motion (Eq. (\ref{tildegammaeq})) is sufficient. At this order we
use the solution for the mode functions Eq. (\ref{modefc}). The
integrals in Eq. (\ref{vev}) can then be performed analytically to
give
\begin{eqnarray}
\langle\Omega|\zeta_n(\B{k}_1)\zeta_n(\B{k}_2)\zeta_n(\B{k}_3)\zeta_n(\B{k}_4)|\Omega\rangle
&=&
(2\pi)^3\delta^3(\B{k}_1+\B{k}_2+\B{k}_3+\B{k}_4)\frac{H^6}{\epsilon^3c_s^3}\frac{1}{\Pi_ik_i^3}\nonumber\\&&
\left[\frac{3}{4}\left(10\Pi+3\lambda\right)\frac{c_s^2}{H^2\epsilon}A_1-\frac{1}{2^6}\left(3\lambda-\frac{H^2\epsilon}{c_s^2}+H^2\epsilon\right)
\frac{1}{H^2\epsilon}A_2-\frac{1}{2^8}\frac{c_s^2-1}{c_s^4}A_3\right],\nonumber\\\label{trizetan}
\end{eqnarray}
where the momentum dependent functions $A_i$ are defined as
\begin{eqnarray}
A_1&=&\frac{\Pi_ik_i^2}{K^5},\nonumber\\
A_2&=&\frac{k_1^2k_2^2(\textbf{k}_3\cdot\textbf{k}_4)}{K^3}\left(1+\frac{3(k_3+k_4)}{K}+\frac{12k_3k_4}{K^2}\right)+\textrm{perm.},
\nonumber \\
A_3&=&\frac{(\textbf{k}_1\cdot\textbf{k}_2)(\textbf{k}_3\cdot\textbf{k}_4)}{K}\left[1+\frac{\sum_{i<j}k_ik_j}{K^2}
+\frac{3k_1k_2k_3k_4}{K^3}
\left(\sum_i\frac{1}{k_i}\right)+12\frac{k_1k_2k_3k_4}{K^4}\right]+\textrm{perm.},
\end{eqnarray}
and ``perm.'' refers to the $24$ permutations of the four momenta.
Note that the quantities of Eq. (\ref{trizetan}) are evaluated at
the moment $\tau_*$ at which the total wave number $K=\sum_{i=1}^4
k_i$ exits the horizon, i.e., when $K{c_s}_*=a_*H_*$. This leading
order result that comes from the 4th order Lagrangian agrees with the result
of Huang and Shiu \cite{Huang:2006eh} \footnote{The full leading order
result for the four point function can be found in the revised version of \cite{Huang:2006eh} which takes into account all the contributions for
the fourth order interaction hamiltonian.} that did their calculation in the uniform
curvature gauge and using a simpler method that is only valid
to calculate the leading order contribution for models with
$c_s\ll 1$. In fact, we can compare our uniform curvature gauge
result (\ref{SSSSphi}) with the result of Huang and Shiu
\cite{Huang:2006eh}. We see that the last terms of the fourth,
sixth and ninth lines of equation (\ref{SSSSphiA}) are exactly the
ones obtained by \cite{Huang:2006eh}, their equation (15), using
the method of just expanding the field lagrangian as in
\cite{Creminelli:2003iq,Gruzinov:2004jx}. For a model with a
general field lagrangian these terms are the ones that give the
leading order contribution for the trispectrum, in the small sound
speed limit, equation (\ref{trizetan}). The contribution coming
from the tensor part will be of next-to-leading order in this
case.

For standard kinetic term inflation, $\Pi=\lambda=0$ and $c_s=1$
and Eq. (\ref{trizetan}) vanishes, the leading order is then given
by the next order in slow roll.

\subsection{\label{subsec:leadingtrispectrumzeta}The leading order trispectrum of $\zeta$}

It is well known that if the slow-roll conditions are satisfied
until the end of inflation and we can ignore gradient terms then
the gauge invariant curvature perturbation $\zeta$ remains
constant on super-horizon scales to all order in perturbation
theory. In this subsection, we will see that this fact greatly
simplifies the relation between the trispectrum of $\zeta$ and
$\zeta_n$.

In the large scales limit, Eq. (\ref{gte}) simplifies to give
\begin{equation}
\zeta=\zeta_n+a\zeta_n^2+\frac{1}{H}\zeta_n\dot{\zeta}_n+\mathcal{O}(\zeta_n^3),\label{gtels}
\end{equation}
where $a$ is defined as
\begin{equation}
a=\frac{\epsilon}{2}+\frac{\ddot{\phi}_0}{2H\dot{\phi}_0}.
\end{equation}
Using the fact that $\dot{\zeta}=0$ on super-horizon scales and
the equation resulting from a time derivative of Eq. (\ref{gtels})
one can show that
\begin{equation}
\dot{\zeta}_n=-\dot{a}\zeta_n^2+\mathcal{O}(\zeta_n^3).\label{zetandot}
\end{equation}
This equation has a simple interpretation. The variable $\zeta_n$
is not constant outside the horizon, only the gauge invariant
quantity $\zeta$ is. This is the reason why the term
$\frac{1}{H}\zeta_n\dot{\zeta}_n$ in the second order gauge
transformation cannot be ignored when one is calculating the
trispectrum of $\zeta$. Substituting Eq. (\ref{zetandot}) in Eq.
(\ref{gte}) and taking the large scale limit we get
\begin{equation}
\zeta=\zeta_n+a\zeta_n^2+ b\zeta_n^3+\cdots \label{ct}
\end{equation}
where $\cdots$ means cubic terms that contain at least one time
derivative of $\zeta_n$ and that will only give a contribution to
the five point function. The variable $b$ is defined as
\begin{eqnarray}
b&=&\frac{\dddot{\phi}_0}{3H^2\dot{\phi}_0}+\frac{\epsilon\ddot{\phi}_0}{H\dot{\phi}_0}+
\frac{\epsilon^2}{3}+\frac{\epsilon\eta}{3}-\frac{\dot{a}}{H}\nonumber\\
&=&-\frac{\dddot{\phi}_0}{6H^2\dot{\phi}_0}+\frac{\epsilon\ddot{\phi}_0^2}{2H\dot{\phi}_0}+\frac{\ddot{\phi}_0^2}{2H^2\dot{\phi}_0^2}
+\frac{\epsilon^2}{3}-\frac{\eta\epsilon}{6}.
\end{eqnarray}

We shall now compare (\ref{ct}) with the result given by the
$\delta N$ formalism
\cite{Starobinsky:1986fxa,Sasaki:1995aw,Lyth:2004gb,Lyth:2005fi}.
In the $\delta N$ approach $\zeta$ is expanded in series in terms
of the field perturbation as
\begin{equation}
\zeta=N'\delta\phi+\frac{1}{2}N''\delta\phi^2+\frac{1}{6}N'''\delta\phi^3+\mathcal{O}(\delta\phi)^4,\label{deltaN}
\end{equation}
where $N$ is the number of e-folds and a prime denotes derivative
with respect to $\phi$. Now comparing Eq. (\ref{ct}) with the
previous equation and observing that
$\zeta_n=-\frac{H}{\dot{\phi}_0}\delta\phi$ we expect
\begin{equation}
\frac{N''}{2}=\frac{H^2}{\dot{\phi}_0^2}a, \quad
\frac{N'''}{6}=-\frac{H^3}{\dot{\phi}_0^3}b.
\end{equation}
We verified that this is indeed the case.

Using Wick's theorem one can now relate the connected part of the
four point correlation function of $\zeta$ with the four point
correlation function of $\zeta_n$ calculated in the previous
section \cite{Byrnes:2006vq,Seery:2006js}. This relation also
involves lower order correlation functions of $\zeta_n$, like the
bispectrum
$\langle\zeta_n(\B{x}_1)\zeta_n(\B{x}_2)\zeta_n(\B{x}_3)\rangle$
(the leading and next-to-leading order in slow roll bispectrum was
previously calculated in \cite{Chen:2006nt}). The relation is
\begin{eqnarray}
\langle\zeta(\B{x}_1)\zeta(\B{x}_2)\zeta(\B{x}_3)\zeta(\B{x}_4)\rangle_c&=&\langle\zeta_n(\B{x}_1)\zeta_n(\B{x}_2)\zeta_n(\B{x}_3)\zeta_n(\B{x}_4)\rangle
\nonumber\\&+&2a\Big[\langle\zeta_n(\B{x}_1)\zeta_n(\B{x}_2)\rangle\langle\zeta_n(\B{x}_1)\zeta_n(\B{x}_3)\zeta_n(\B{x}_4)\rangle
+\langle\zeta_n(\B{x}_1)\zeta_n(\B{x}_3)\rangle\langle\zeta_n(\B{x}_1)\zeta_n(\B{x}_2)\zeta_n(\B{x}_4)\rangle
\nonumber\\&&\quad+\langle\zeta_n(\B{x}_1)\zeta_n(\B{x}_4)\rangle\langle\zeta_n(\B{x}_1)\zeta_n(\B{x}_2)\zeta_n(\B{x}_3)\rangle
+3\;perm \Big]
\nonumber\\&+&4a^2\Big[\langle\zeta_n(\B{x}_1)\zeta_n(\B{x}_2)\rangle\langle\zeta_n(\B{x}_1)\zeta_n(\B{x}_3)\rangle\langle\zeta_n(\B{x}_2)\zeta_n(\B{x}_4)\rangle
\nonumber\\&&\quad+\langle\zeta_n(\B{x}_1)\zeta_n(\B{x}_2)\rangle\langle\zeta_n(\B{x}_1)\zeta_n(\B{x}_4)\rangle\langle\zeta_n(\B{x}_2)\zeta_n(\B{x}_3)\rangle
+ 5\;perm\Big]
\nonumber\\&+&6b\Big[\langle\zeta_n(\B{x}_1)\zeta_n(\B{x}_2)\rangle\langle\zeta_n(\B{x}_1)\zeta_n(\B{x}_3)\rangle\langle\zeta_n(\B{x}_1)\zeta_n(\B{x}_4)\rangle
+ 3\;perm \Big] + \mathcal{O}(P^\zeta_k)^4,
\end{eqnarray}
where ``perm" means the other permutations of the preceding terms
and $\mathcal{O}(P^\zeta_k)^4$ denotes terms that are suppressed
by successive powers of the power spectrum.  Now, one can easily
see that at leading order in slow roll the trispectrum for $\zeta$
and $\zeta_n$ are equal, this is because the constants $a$ and $b$
are slow-roll suppressed. These terms will only contribute to the
next-to-leading order corrections.

\subsection{\label{subsec:nextobleadingcorrections}The next-to-leading order
corrections for the trispectrum}

In subsection \ref{subsec:leadingtrispectrumzetan}, we showed that
for standard kinetic term inflation the leading order result,
equation (\ref{trizetan}), vanishes and in fact in this case the
leading order of the trispectrum of $\zeta_n$ is of order
$\epsilon^{-2}$ (the next-to-leading order is the leading order).
To obtain these leading order contributions it is easier to
perform the calculation using the uniform curvature gauge action
Eq. (\ref{SSSSphiA})-(\ref{GSSphi}). Eq. (\ref{SSSSphiA}) vanishes
exactly for standard kinetic term inflation. The action
(\ref{SSSSphiB}) is exact in the slow-roll approximation but it is
instructive to determine the slow-roll order of the different
terms. One can see that the leading order contribution comes from
terms of order $\mathcal{O}(\epsilon^0)$, as pointed out in
\cite{Seery:2006vu}. If we take $P(X,\phi)=X-V(\phi)$ then the
leading order (in slow roll) source of Eq. (\ref{tildegammaeq})
will be of order $\mathcal{O}(\epsilon^0)$. We therefore don't
expect $\tilde{\gamma}_{ij}$ to be slow-roll suppressed and the
actions (\ref{GGphi}), (\ref{GSSphi}) will contain unsuppressed
terms of the same order as the leading order term of the action
(\ref{SSSSphiB}). These tensor contributions were absent in the
analysis of \cite{Seery:2006vu} and we have shown that they are of
the same order as the fourth order action considered in
\cite{Seery:2006vu}, our Eq. (\ref{SSSSphiB}). It is still an open
question how these new contributions will change the trispectrum
result of Seery \emph{et al.}.

For the general lagrangian case, the leading order trispectrum was
given in the previous subsection and in \cite{Huang:2006eh}.
Contrary to the method of \cite{Huang:2006eh}, our method of
obtaining the fourth order action (\ref{SSSSphi}) does not rely on
any approximation and therefore the action (\ref{SSSSphi}) is
valid to all orders in slow roll and in the sound speed expansion
and it can be used to study the next-to-leading order corrections.
Depending on the momentum shape of these next-to-leading terms
they might become big enough to be observed in the next generation
of experiments. A similar argument applies for the next-to-leading
order corrections for the bispectrum, as it was shown in
\cite{Chen:2006nt}. For example, for DBI inflation,
\cite{Huang:2006eh} showed that the leading order non-gaussianity
parameters $\tau_{NL}$ scales like $\tau_{NL}\sim0.1/c_s^4$ (for a
specific momentum configuration) and $f_{NL}\sim1/c_s^2$. They
argue that if $c_s\sim0.1$ then $f_{NL}$ is still inside the value
range allowed by observations but $\tau_{NL}\sim 10^3$ could be
detected with the Planck satellite CMBR experiment. Therefore,
assuming that the slow-roll parameter $\epsilon$ is of order
$\epsilon\sim0.01$ (at horizon crossing) these next-to-leading
order corrections for the trispectrum could possible be observed
with the Planck satellite. A more careful and systematic study of
the momentum dependence of these new terms is required and it is
left for future work.

\section{\label{sec:conclusion}Conclusion}

We have computed the fourth order action for scalar and second
order tensor perturbations in the comoving gauge. Our result is
exact in the slow-roll (SR) expansion but practically it is useful
to study the SR suppression of the different terms. We were able
to show that after many integrations by parts the unsuppressed
terms contained in the previous action can be reduced to total
derivatives terms plus corrections that are SR suppressed. The
resulting action has the correct order in SR. It is suppressed by
$\epsilon$ as it should be, because in pure de Sitter space the
curvature perturbation is a pure gauge mode. An important lesson
from our work is that in order to obtain the correct SR order for
the action, the second order tensor perturbations cannot be
ignored as assumed in previous works \cite{Seery:2006vu} and
\cite{Jarnhus:2007ia}. We found the explicit form of these tensor
perturbations in the comoving gauge by using the gauge
transformations from the uniform curvature gauge. Fortunately, for
a general inflation model like (\ref{action}), we showed that we
do not need to solve the equations of motion for the tensor
perturbations if we are interested in calculating the leading
order trispectrum. However, to calculate the next-to-leading order
corrections to that result, or to calculate the leading order
trispectrum for standard kinetic term inflation, we do need to
solve explicitly the equations of motion for the tensor
perturbations. This will be left for future work.

Using the comoving gauge action we have calculated the leading
order in SR trispectrum of $\zeta$. We compared our result with
the result of \cite{Huang:2006eh}, obtained using the uniform
curvature gauge, and we found an agreement.

For the uniform curvature gauge action, that is also exact in the
SR expansion, we identified the terms that will contribute to the
next-to-leading order corrections to the trispectrum. We pointed
out that depending on the model and on the momentum configuration,
some of these corrections might be observable with the Planck
satellite. After taking particular limits, the previous action
nicely reduces to previously know results \cite{Seery:2006vu},
\cite{Huang:2006eh} with the caveat that the above mentioned works
ignore tensor contributions.

Finally we have obtained the relations between the trispectrum of
$\zeta$ and $\delta\phi$ (on large scales) using the third order
gauge transformations and compared the result with the $\delta N$
formalism.

To conclude, we have provided the necessary equations (fourth
order action and the relation between $\zeta$ and $\delta\phi$) to
calculate the trispectrum for a fairly general model of inflation
that are also valid for models where SR is temporarily
interrupted, i.e., around a ``step" in the inflaton's lagrangian
\cite{Hailu:2006uj}. In this case, it is impossible to apply the
$\delta N$ approach and it is required to evaluate the n-point
functions numerically \cite{Chen:2006xjb} (see
\cite{Tanaka:2007gh} for a different approach). We leave this more
practical application of our results for future work.

\begin{acknowledgments}
We thank Misao Sasaki for useful discussions. FA is supported by
``Funda\c{c}\~{a}o para a Ci\^{e}ncia e Tecnologia (Portugal)",
with the fellowship's reference number: SFRH/BD/18116/2004. KK is
supported by STFC.
\end{acknowledgments}

\appendix
\section{\label{GT}Gauge transformations up to second order}

In this Appendix we will find the change of variables that one
needs to perform to go from the uniform curvature gauge
(\ref{deltaphigauge}) to the comoving gauge (\ref{zetagauge}). A
similar result can be found in \cite{Maldacena:2002vr}. In order
to go from the gauge (\ref{deltaphigauge}) where the field
fluctuation is not zero to the gauge (\ref{zetagauge}) where
$\delta\phi=0$ we need a change of variables that satisfy
$\phi(t+T(t))+\delta\phi(t+T(t))=\phi(t)$.

At first order in perturbation theory we only need to do a time
reparametrization. Let $t$ and $\tilde{t}$ be the time coordinates
in the gauges (\ref{zetagauge}) and (\ref{deltaphigauge})
respectively. The time reparametrization is $\tilde{t}=t+T$. At
first order
\begin{eqnarray}
T=-\frac{\delta\phi}{\dot{\phi_0}}=\frac{\zeta}{H}, \quad
\zeta=-\frac{H}{\dot{\phi_0}}\delta\phi.
\end{eqnarray}
At second order the time reparametrization is
\begin{equation}
T=-\frac{\delta\phi}{\dot{\phi_0}}-\frac{\ddot{\phi_0}\delta\phi^2}{2\dot{\phi_0}^3}+\frac{\dot{\delta\phi}\delta\phi}{\dot{\phi}^2}.
\end{equation}
At this order we also need to perform a spatial reparametrization
given by $\tilde{x}^i=x^i+\epsilon^i(x,t)$, where $\epsilon^i$ is
of second order in the perturbations. The metric in the gauge
(\ref{zetagauge}) becomes
\begin{equation}
h_{ij}=-\frac{\partial T}{\partial x^i}\frac{\partial T}{\partial
x^j}+N_j^{(1)}\frac{\partial T}{\partial
x^i}+N_i^{(1)}\frac{\partial T}{\partial
x^j}+a^2T\dot{\tilde{\gamma}}_{ij}+a^2\left(\frac{\partial
\epsilon_j}{\partial x^i}+\frac{\partial \epsilon_i}{\partial
x^j}\right)
+a^2e^{2HT+\dot{H}T^2}\left(\delta_{ij}+\tilde{\gamma}_{ij}(t)+\frac{1}{2}\tilde{\gamma}_{ik}\tilde{\gamma}^k_j\right),
\label{CG}
\end{equation}
where $N_i^{(1)}$ is the first order shift vector in the gauge
(\ref{deltaphigauge}). If the vector $\epsilon^i$ obeys the
equation
\begin{equation}
a^{-2}\delta h_{ij}+\frac{\partial \epsilon_j}{\partial
x^i}+\frac{\partial \epsilon_i}{\partial
x^j}=2\beta\delta_{ij}+\mu_{ij},\label{CG2}
\end{equation}
with $\mu_{ij}$ being a transverse and traceless tensor and
$\delta h_{ij}$ being defined as the first four terms of Eq.
(\ref{CG}), then the gauge transformation equations are given by
\begin{eqnarray}
\zeta&=&HT+\frac{\dot{H}T^2}{2}+\beta,\nonumber\\
\gamma_{ij}&=&\tilde{\gamma}_{ij}(t)+\mu_{ij}.
\end{eqnarray}
To obtain the quantities $\beta$ and $\mu_{ij}$ it proves to be
useful to decompose $\epsilon^i$ in
$\epsilon^i=\partial^i\tilde{\epsilon}+\epsilon_t^i$ with
$\partial_i\epsilon_t^i=0$. After a few mathematical manipulations
of equation (\ref{CG2}) one can obtain
\begin{equation}
\beta=\frac{a^{-2}}{4}\left(\delta
h_i^i-\partial^{-2}\partial^i\partial^j\delta
h_{ij}\right),\label{beta}
\end{equation}
\begin{equation}
\mu_{ij}=a^{-2}\left(\delta h_{ij}-\frac{1}{2}\delta_{ij}\delta
h_k^k-\partial^{-2}\partial_i\partial^k\delta
h_{kj}-\partial^{-2}\partial_j\partial^k\delta h_{ki}
+\frac{1}{2}\delta_{ij}\partial^{-2}\partial^l\partial^k\delta
h_{lk}+\frac{1}{2}\partial^{-2}\partial_i\partial_j\delta h_k^k
+\frac{1}{2}\partial^{-4}\partial_i\partial_j\partial^l\partial^k\delta
h_{lk} \right),
\end{equation}
where $\delta h_{ij}$ can be written explicitly as
\begin{equation}
\delta
h_{ij}=-\frac{1}{H^2}\partial_i\zeta_n\partial_j\zeta_n+\frac{1}{H}\left(\partial_i\zeta_n\partial_j\psi_1+\partial_j\zeta_n\partial_i\psi_1\right)
+\frac{a^2}{H}\zeta_n\dot{\tilde{\gamma}}_{ij},
\end{equation}
where $\psi_1$ is from the uniform curvature gauge and we have
used the variable $\zeta_n$ introduced before, Eq. (\ref{zetan}).
As $\tilde{\gamma}_{ij}$ is of second order now, the term
$\zeta_n\dot{\tilde{\gamma}}_{ij}$ is of third order. We kept it
in Eq. (\ref{CG}) for the sake of comparison with the result of
\cite{Maldacena:2002vr}. For $\zeta$ we have
\begin{equation}
\zeta=\zeta_n+\frac{\epsilon}{2}\zeta_n^2+\frac{\ddot{\phi}_0}{2\dot{\phi}_0H}\zeta_n^2+\frac{1}{H}\zeta_n\dot{\zeta_n}+\beta.
\label{GTzeta}
\end{equation}

\section{\label{TTpart}Extraction of TT part of a tensor}

Let $T_{ij}$ to be a given 3D symmetric tensor, as it is the case
for the source of equation (\ref{tildegammaeq}). Then it can be
decomposed into a trace part and a traceless part as
\begin{equation}
T_{ij}=\frac{T}{3}\delta_{ij}+\tilde{T}_{ij}.
\end{equation}
The traceless part (5 degrees of freedom) can be written like
\begin{equation}
\tilde{T}_{ij}=D_{ij}\chi+\partial_i\chi_j+\partial_j\chi_i+\chi_{ij},
\end{equation}
with
$D_{ij}\equiv\partial_i\partial_j-\frac{1}{3}\delta_{ij}\partial^2$,
$\partial^i\chi_i=0$ and $\partial^i\chi_{ij}=0=\chi_i^i$, where
indices are raised by $\delta_{ij}$. The equation
$\partial^iT_{ij}=\frac{1}{3}\partial_jT+\frac{2}{3}\partial_j\partial^2\chi+\partial^2\chi_j$
can be solved using a similar method as the one we used to solve
the second order momentum constraint previously. We then find
\begin{equation}
\chi=\frac{3}{2}\partial^{-4}\partial^jF_j, \quad
\chi_j=\partial^{-2}F_j-\partial_j\partial^i\partial^{-4}F_i,
\end{equation}
where $F_i\equiv\partial^jT_{ij}-\frac{1}{3}\partial_iT$. And
\begin{equation}
\chi_{ij}=T_{ij}-\frac{T}{3}\delta_{ij}-D_{ij}\chi-\partial_i\chi_j-\partial_j\chi_i.
\label{TT}
\end{equation}
In conclusion, given a tensor $T_{ij}$, Eq. (\ref{TT}) defines its
transverse and traceless part.

Let us see how this works at the action level for the particular
case of the tensor perturbations described in the main text. In
the action (\ref{GSSphi}), the source for $\tilde{\gamma}_{ij}$ is
of the form
\begin{equation}
S=\int dtd^3x\tilde{\gamma}^{ij}T_{ij}.
\end{equation}
with $T_{ij}$ being quadratic in $\delta\phi$. We can see that
because $\tilde{\gamma}_{ij}$ is transverse and traceless we are
allowed to replace $T_{ij}$ in the previous action with
$\chi_{ij}$ defined in (\ref{TT}). If we calculate the equations
of motion by varying the resulting action we get as a source
$\chi_{ij}$ and not simply $T_{ij}$ (see Eq.
(\ref{tildegammaeq})), ensuring that both sides of the equations
of motion are transverse and traceless.


\begin{thebibliography}{10}

\bibitem{Spergel:2006hy}
WMAP, D.~N. Spergel {\em et~al.},
\newblock Astrophys. J. Suppl. {\bf 170}, 377 (2007), astro-ph/0603449.

\bibitem{Komatsu:2001rj}
E.~Komatsu and D.~N. Spergel,
\newblock Phys. Rev. {\bf D63}, 063002 (2001), astro-ph/0005036.

\bibitem{Boubekeur:2005fj}
L.~Boubekeur and D.~H. Lyth,
\newblock Phys. Rev. {\bf D73}, 021301 (2006), astro-ph/0504046.

\bibitem{Alabidi:2005qi}
L.~Alabidi and D.~H. Lyth,
\newblock JCAP {\bf 0605}, 016 (2006), astro-ph/0510441.

\bibitem{Kogo:2006kh}
N.~Kogo and E.~Komatsu,
\newblock Phys. Rev. {\bf D73}, 083007 (2006), astro-ph/0602099.

\bibitem{Babich:2004gb}
D.~Babich, P.~Creminelli, and M.~Zaldarriaga,
\newblock JCAP {\bf 0408}, 009 (2004), astro-ph/0405356.

\bibitem{Maldacena:2002vr}
J.~M. Maldacena,
\newblock JHEP {\bf 05}, 013 (2003), astro-ph/0210603.

\bibitem{Seery:2005wm}
D.~Seery and J.~E. Lidsey,
\newblock JCAP {\bf 0506}, 003 (2005), astro-ph/0503692.

\bibitem{Seery:2005gb}
D.~Seery and J.~E. Lidsey,
\newblock JCAP {\bf 0509}, 011 (2005), astro-ph/0506056.

\bibitem{Chen:2006nt}
X.~Chen, M.-x. Huang, S.~Kachru, and G.~Shiu,
\newblock JCAP {\bf 0701}, 002 (2007), hep-th/0605045.

\bibitem{Seery:2006vu}
D.~Seery, J.~E. Lidsey, and M.~S. Sloth,
\newblock JCAP {\bf 0701}, 027 (2007), astro-ph/0610210.

\bibitem{Alishahiha:2004eh}
M.~Alishahiha, E.~Silverstein, and D.~Tong,
\newblock Phys. Rev. {\bf D70}, 123505 (2004), hep-th/0404084.

\bibitem{Huang:2006eh}
M.-x. Huang and G.~Shiu,
\newblock Phys. Rev. {\bf D74}, 121301 (2006), hep-th/0610235.

\bibitem{Gruzinov:2004jx}
A.~Gruzinov,
\newblock Phys. Rev. {\bf D71}, 027301 (2005), astro-ph/0406129.

\bibitem{Jarnhus:2007ia}
P.~R. Jarnhus and M.~S. Sloth,
\newblock (2007), arXiv:0709.2708 [hep-th].

\bibitem{Chen:2006xjb}
X.~Chen, R.~Easther, and E.~A. Lim,
\newblock JCAP {\bf 0706}, 023 (2007), astro-ph/0611645.

\bibitem{Chen:2008wn}
X.~Chen, R.~Easther, and E.~A. Lim,
\newblock (2008), arXiv:0801.3295 [astro-ph].

\bibitem{Hailu:2006uj}
G.~Hailu and S.~H.~H. Tye,
\newblock JHEP {\bf 08}, 009 (2007), hep-th/0611353.

\bibitem{Silverstein:2003hf}
E.~Silverstein and D.~Tong,
\newblock Phys. Rev. {\bf D70}, 103505 (2004), hep-th/0310221.

\bibitem{ArmendarizPicon:1999rj}
C.~Armendariz-Picon, T.~Damour, and V.~F. Mukhanov,
\newblock Phys. Lett. {\bf B458}, 209 (1999), hep-th/9904075.

\bibitem{Garriga:1999vw}
J.~Garriga and V.~F. Mukhanov,
\newblock Phys. Lett. {\bf B458}, 219 (1999), hep-th/9904176.

\bibitem{Acquaviva:2002ud}
V.~Acquaviva, N.~Bartolo, S.~Matarrese, and A.~Riotto,
\newblock Nucl. Phys. {\bf B667}, 119 (2003), astro-ph/0209156.

\bibitem{Arnowitt:1960es}
R.~Arnowitt, S.~Deser, and C.~W. Misner,
\newblock Phys. Rev. {\bf 117}, 1595 (1960).

\bibitem{Ananda:2006af}
K.~N. Ananda, C.~Clarkson, and D.~Wands,
\newblock Phys. Rev. {\bf D75}, 123518 (2007), gr-qc/0612013.

\bibitem{Osano:2006ew}
B.~Osano, C.~Pitrou, P.~Dunsby, J.-P. Uzan, and C.~Clarkson,
\newblock JCAP {\bf 0704}, 003 (2007), gr-qc/0612108.

\bibitem{Weinberg:2005vy}
S.~Weinberg,
\newblock Phys. Rev. {\bf D72}, 043514 (2005), hep-th/0506236.

\bibitem{Creminelli:2003iq}
P.~Creminelli,
\newblock JCAP {\bf 0310}, 003 (2003), astro-ph/0306122.

\bibitem{Starobinsky:1986fxa}
A.~A. Starobinsky,
\newblock JETP Lett. {\bf 42}, 152 (1985).

\bibitem{Sasaki:1995aw}
M.~Sasaki and E.~D. Stewart,
\newblock Prog. Theor. Phys. {\bf 95}, 71 (1996), astro-ph/9507001.

\bibitem{Lyth:2004gb}
D.~H. Lyth, K.~A. Malik, and M.~Sasaki,
\newblock JCAP {\bf 0505}, 004 (2005), astro-ph/0411220.

\bibitem{Lyth:2005fi}
D.~H. Lyth and Y.~Rodriguez,
\newblock Phys. Rev. Lett. {\bf 95}, 121302 (2005), astro-ph/0504045.

\bibitem{Byrnes:2006vq}
C.~T. Byrnes, M.~Sasaki, and D.~Wands,
\newblock Phys. Rev. {\bf D74}, 123519 (2006), astro-ph/0611075.

\bibitem{Seery:2006js}
D.~Seery and J.~E. Lidsey,
\newblock JCAP {\bf 0701}, 008 (2007), astro-ph/0611034.

\bibitem{Tanaka:2007gh}
Y.~Tanaka and M.~Sasaki,
\newblock (2007), arXiv:0706.0678 [gr-qc].

\end{thebibliography}

\end{document}